\documentclass[11pt,reqno,oneside]{amsart}
\usepackage{amssymb}
\usepackage[reqno]{amsmath}

\begin{document}

\begin{flushright}
\baselineskip=12pt
CERN--TH/97--167\\
%\tt gr-qc/97?????
\end{flushright}

\title{Issues in Quantum--Geometric Propagation}
\author{M.A. Clayton}
\address{CERN--Theory Division, CH--1211 Geneva 23, Switzerland}
\email{Michael.A.Clayton@cern.ch}
\date{\today}
%\thanks{CERN--??}

\begin{abstract}
A discussion of relativistic quantum--geometric mechanics on phase space and its generalisation to the propagation of free, massive, quantum--geometric scalar fields on curved spacetimes is given.
It is shown that in an arbitrary coordinate system and frame of reference in a flat spacetime, the resulting propagator is necessarily the same as derived in the standard Minkowski coordinates up to a Lorentz boost acting on the momentum content of the field, which is therefore seen to play the role of Bogolubov transformations in this formalism.
These results are explicitly demonstrated in the context of a Milne universe.
\end{abstract}
\maketitle

\section{Introduction}
\label{sect:Intro}

That a fundamental scale should exist in the `ultimate' theory of nature seems by now to be fairly well accepted, and it is therefore of interest to examine a program that incorporates such a scale into physics from the outset~\cite{Prugovecki:1984,Prugovecki:1992,Prugovecki:1995}.
That the program involves a modification of non--relativistic quantum mechanics (without disturbing its agreement with experiment) is overshadowed by the improved measurement--theoretic scenario that emerges, as well as the extension to a relativistic quantum mechanical picture in both flat and curved spacetimes; leading to a quantum gravitational scenario.
The twofold purpose of this manuscript is to give an overview of the basic ideas and results of scalar Quantum--Geometric (QG) field theory on a curved spacetime, as well as to clarify QG field propagation in flat spacetime using a Milne universe for illustrative purposes.
Even in such a simple spacetime model the conventional field quantisation is not without it's ambiguities~\cite{Arcuri+Svaiter+Svaiter:1994,Tanaka+Sasaki:1996}.

That the conventional treatment of local quantum fields, or the resulting formal `perturbative' series, is not adequate follows from attempts to construct the simplest scalar quantum field theory.
These results indicate that, at best, the current perturbative treatment is not in accord with the these rigorous results~\cite[Section 15.1.6]{Fernandez+Frolich+Sokal:1992} and it is felt that ultimately the model will turn out to be trivial~\cite[Section 21.6]{Glimm+Jaffe:1987}.
Despite this, it is by now also well--established that the formal perturbation theory that results from considering such models must have some physical validity, due to the high accuracy of its agreement with experimental results~\cite{Kinoshita:1990,Donoghue+Golowich+Holstein:1992}.
One of the remarkable results from QG field theory is that the series that results from considering the analogous QG field theory models formally reproduces the conventional series term by term in the sharp point limit (\textit{i.e.}, when the fundamental length scale is taken to zero).
Thus although the question of triviality of such models has not been rigorously answered, the formal agreement of QG models with experiment is established, even at such high accuracies.

Furthermore, the naive extension of local quantum fields to curved spacetimes provides a ``grossly inadequate foundation for the theory~\cite{DeWitt:1975}'' as it requires an (approximate) timelike Killing vector for its formulation~\cite{Kuchar:1988}.
Although QG field propagation is free of these criticisms~\cite{Prugovecki:1994}, as we shall see there are still issues to be resolved.
However it is interesting that the consistent incorporation of a quantum analogue of the strong equivalence principle into the formalism implies that the violation of local energy--momentum that leads to \textit{ex--nihilo} particle production in non--inertial frames of reference in the conventional field theory \textit{cannot} occur.

The plan of the manuscript is straightforward; we begin by giving an overview of Minkowski space relativistic stochastic quantum mechanics and conclude the general discussion with the geometrization of this program and some results from generic flat spacetime models.
Following this, we make some of these results more concrete by considering a Milne universe.

\section{Background}
\label{sect:background}

Here we will describe that part of relativistic QG mechanics that is necessary background to consider the evolution of free QG scalar fields in a classical, globally hyperbolic spacetime; further details on interacting quantum fields and gauge fields has appeared elsewhere~\cite{Prugovecki:1995}.
Although we shall discuss relativistic QG fields exclusively, The construction proceeds in an analogous manner in the case of non-relativistic fields~\cite{Ali:1985}, and may be reached through a group contraction of the Poincar\'{e} group of the relativistic QG formalism to the Galilei group of the non-relativistic QG formalism~\cite{Prugovecki:1984}.

Loosely speaking, the wave function $\psi(x)$ of the conventional (local) theory which represents the amplitude to find the system localised about a point $x$, is replaced in the QG formalism by a wave function $\psi(q,p)$ which contains information about the spatial localisation as well as the velocity of the system.
However whereas the local wave function of the former (and its extension to a local field operator) is intended to allow for arbitrarily sharply localised position measurements of the state, the outcome of measurements on the latter state may only be interpreted in a stochastic sense.
That is, instead of the practically unrealizable limit of measurements of the field at a point~\cite{Bohr+Rosenfeld:1950}, the outcome of the sharpest possible measurement of the position and momentum of a QG field would indicate that the state is stochastically localised about $q$ with a (stochastic) average momentum $p$.
The distribution of possible values about this outcome is related to the fundamental scale through a resolution generator; a necessary byproduct of the reduction of the Poincar\'{e} group action on phase space, also leading to a conserved probability current with a positive--definite timelike component.

As we shall see, emphasis is placed on a local vacuum which represents (roughly speaking) a micro-detector at stochastic rest, stochastically localised about the spacetime point in question.
Due to the fact that the formalism does \textit{not} rely on the existence of a global vacuum for its definition, the geometrization of the flat spacetime theory in Section~\ref{sect:geometrization} survives the passage to curved spacetimes.

\subsection{Relativistic Quantum--Geometric Mechanics}
\label{sect:RQM}

To begin we must introduce some notation.
The action of the Poincar\'{e} group $\mathrm{ISO}_0(1,3)$ on the phase space manifold $\Gamma=\mathbb{M}^4\times V_m^+\subset\mathbb{R}^8$ of pairs $(q,p)$ with $q\in\mathbb{M}^4=(\mathbb{R}^4,\eta)$ from Minkowski space ($\eta=\mathrm{diag}(1,-1,-1,-1)$) and $p\in V_m^+=\{p\in\mathbb{R}^4\mid p^2=m^2, p^0>0\}$ from the positive mass hyperboloid is
\begin{subequations}
\begin{equation}
(\Lambda,b)\star(q,p)=(\Lambda q+b,\Lambda p),\quad
\Lambda\in\mathrm{SO}(1,3),\quad b\in \mathbb{R}^4,
\end{equation}
with Poincar\'{e} group product and inverse given by
\begin{equation}
(\Lambda_2,b_2)\star(\Lambda_1,b_1)=(\Lambda_2\Lambda_1,\Lambda_2b_1+b_2),\quad
(\Lambda,b)^{-1}=(\Lambda^{-1},-\Lambda^{-1}b).
\end{equation}
\end{subequations}
Here we will use the convention that, for example, $q$ represents the vector components $q^A$ ($A,B,\ldots\in\{0,1,2,3\}$), $\Lambda q$ the product ${\Lambda^A}_Bq^B$ (so we will not indicate the contraction explicitly), and the product of two vectors is written as $p\cdot q:=p^Aq^B\eta_{AB}$.
From the definition of the Lorentz matrices $\eta^{AB}={\Lambda^A}_C{\Lambda^B}_D\eta^{CD}$ we have that $p\cdot \Lambda q=\Lambda^{-1}p\cdot q$ where ${(\Lambda^{-1})^A}_B=\eta^{AC}\eta_{BD}{\Lambda^D}_C$.

Defining the product $\Sigma_m=\sigma\times V_m^+\subset\mathbb{M}^4\times V_m^+$ of a maximal spacelike hypersurface $\sigma$ in $\mathbb{M}^4$ and the forward mass hyperboloid, one introduces the Hilbert space of square--integrable functions on these surfaces $L^2(\Sigma_m)$ with inner product
\begin{equation}\label{eq:inner product}
\langle \phi_1|\phi_2\rangle =\int_{\Sigma_m}\phi_1^*(q,p)\phi_2(q,p) 
d\Sigma_m(q,p),
\end{equation}
where the Poincar\'{e} invariant measure is given by~\cite{Prugovecki:1984}
\begin{subequations}
\begin{gather}
d\Sigma_m(q,p):=2p\cdot d\sigma(q)d\Omega_m(p),\\
d\Omega_m(p)=\delta(p^2-m^2)d^4p=
\frac{d^3\vec{p}}{2p^0}\Big\rvert_{p^0=\sqrt{\vec{p}^2+m^2}},
\end{gather}
\end{subequations}
and $d\sigma^A(q)$ is the volume element of the hypersurface $\sigma\in\mathbb{M}^4$.
Note that $d\Sigma_m$ is normalised so that on a spatial hyper-plane defined by $q^0=0$ it reduces to $d^3\vec{q}\,d^3\vec{p}$.

The induced spin zero representation of $\mathrm{ISO}_0(1,3)$ on $\Gamma$ 
\begin{equation}\label{eq:poincare}
U(\Lambda,b):\phi(q,p)\rightarrow
\bigl(U(\Lambda,b)\phi\bigr)(q,p)=\phi\bigl((\Lambda,b)^{-1}\star(q,p)\bigr)
=\phi\bigl(\Lambda^{-1}(q-b),\Lambda^{-1}p\bigr),
\end{equation}
was found (in contrast to the usual representations on momentum or configuration space~\cite[Chapter 7]{Bogolubov+:1990}) to be highly reducible~\cite[Section 3.4]{Prugovecki:1992}.
Each irreducible sub-representation defines a subspace of $L^2(\Sigma_m)$ of functions of the form
\begin{equation}\label{eq:subirrep}
\phi(q,p)
=\int_{V_m^+}\mathrm{e}^{-iq\cdot k}
\tilde{\eta}\Bigl(\frac{p\cdot k}{m}\Bigr)\tilde{\phi}(k)d\Omega_m(k)
=\int_{V_m^+}
\tilde{\eta}^*_{q,p}(k)\tilde{\phi}(k)d\Omega_m(k),
\end{equation}
defining the Hilbert space $\mathbf{P}L^2(\Sigma_m)$.
Here we have assumed that the irreducible sub-representations are characterised by a real, rotationally invariant resolution generator $\tilde{\eta}$, which in this work will be chosen as
\begin{equation}
\tilde{\eta}(k)=\sqrt{\frac{m^3}{\tilde{Z}_{l,m}}}\mathrm{e}^{-lk^0},\quad k\in V_m^+,
\end{equation}
which is a minimum uncertainty state, and emerges as the unique ground state wave function of the reciprocally invariant quantum metric operator~\cite[Section 4.5]{Prugovecki:1984}.
(Note that all of the relations given here may either be viewed in Planck units~\cite[Appendix F]{Wald:1984} where all quantities are dimensionless, or in units where $\hbar=c=1$ so that $l$ and $m^{-1}$ retain the dimensions of a length.
In the former case it is still useful to retain the parameter $l$ in order to examine the sharp point limit $l\rightarrow 0$.)

Considering the infinitesimal form of the Poincar\'{e} transformation in~\eqref{eq:poincare}: $(\Lambda,b)\approx (\mathbf{1},0)+(\kappa,\beta)$, where $\kappa=\tfrac{1}{2}{\kappa^A}_B{M^B}_A$ and $\beta=\beta_AP^A$, we find that the generators $\{{M^A}_B,P^A\}$ of the Lie algebra $\mathfrak{iso}(1,3)$ acting on $\mathbf{P}L^2(\Sigma_m)$ as
\begin{equation}
\delta\phi(q,p)
=\bigl(\beta_AP^A+\tfrac{1}{2}{\kappa^A}_B{M^B}_A\bigr)\phi(q,p)
=i\bigl(\beta_A\tilde{P}^A
+\tfrac{1}{2}{\kappa^A}_B{{ }\tilde{M}^B}_A\bigr)\phi(q,p)
\end{equation}
satisfy the Lie algebra of $\mathrm{ISO}(1,3)$
\begin{subequations}\label{eq:generators}
\begin{gather}
[P_A,P_B]=0,\quad
[M_{AB},P_C]=\eta_{AC}P_B-\eta_{BC}P_A,\\
[M_{AB},M_{CD}]=
\eta_{AC}M_{BD}-\eta_{AD}M_{BC}
-\eta_{BC}M_{AD}+\eta_{BD}M_{AC},
\end{gather}
\end{subequations}
and are defined in terms of the eight operators~\cite[Section 4.4]{Prugovecki:1995}
\begin{subequations}
\begin{equation}
\tilde{P}_A:=i\partial_{q^A},\quad 
\tilde{Q}_A:=q_A-i\partial_{p^A},
\end{equation}
as
\begin{equation}
P_A=i\tilde{P}_A,\quad
M_{AB}=i\tilde{M}_{AB}=:i(\tilde{Q}_A\tilde{P}_B-\tilde{Q}_B\tilde{P}_A).
\end{equation}
\end{subequations}
Furthermore, the operators $\tilde{P}_A$ and $\tilde{Q}_A$ satisfy the relativistic commutation relations
\begin{equation}
[\tilde{Q}_A,\tilde{P}_B]=-i\eta_{AB},\quad
[\tilde{Q}_A,\tilde{Q}_B]=[\tilde{P}_A,\tilde{P}_B]=0,
\end{equation}
which is related to the relativistic generalisation of the Heisenberg uncertainty relations~\cite[Section~3.8]{Prugovecki:1995}.

The collection of
\begin{equation}\label{eq:tilde frames}
\tilde{\eta}_{q,p}(k)
:=\bigl(\tilde{U}(\Lambda_{p/m},q)\tilde{\eta}\bigr)(k)
=\sqrt{\frac{m^3}{\tilde{Z}_{l,m}}}
\mathrm{e}^{ik\cdot q-\frac{l}{m}k\cdot p}
=\sqrt{\frac{m^3}{\tilde{Z}_{l,m}}}
\mathrm{e}^{ik\cdot \zeta^*},\quad
(q,p)\in \Gamma,
\end{equation}
($\Lambda_{p/m}$ is the boost to velocity $p/m$, and we have introduced the complex variables $\zeta:=q-i\tfrac{l}{m}p$) generated by the action of the Poincar\'{e} group on the resolution generator, allows the definition of the projection of $\phi\in L^2(\Sigma_m)$ onto $\mathbf{P}L^2(\Sigma_m)$ via the kernel
\begin{equation}\label{eq:kernel}
K(q_2,p_2;q_1,p_1)
:=\int_{V_m^+}d\Omega_m(k)\tilde{\eta}^*_{q_2,p_2}(k)\tilde{\eta}_{q_1,p_1}(k)
=\frac{m^3}{\tilde{Z}_{l,m}}
\int_{V_m^+}d\Omega_m(k)\mathrm{e}^{-ik\cdot(\zeta_2-\zeta_1^*)}, 
\end{equation}
as
\begin{equation}\label{eq:scalar}
\phi(q^\prime,p^\prime)
=\int_{\Sigma_m}K(q^\prime,p^\prime;q,p)\phi(q,p)d\Sigma_m(q,p).
\end{equation}
Choosing the normalisation constant $\tilde{Z}_{l,m}$ as~\cite{Prugovecki:1995}
\begin{equation}
\tilde{Z}_{l,m}=(2\pi)^4m^5\frac{K_2(2lm)}{2lm},
\end{equation}
implies that $\langle\phi_1|\phi_2\rangle=\langle\tilde{\phi}_1|\tilde{\phi}_2\rangle$, and the kernel in~\eqref{eq:kernel} acts as a reproducing kernel on $\mathbf{P}L^2(\Sigma_m)$~\cite[Section 3.6]{Prugovecki:1992}
\begin{equation}\label{eq:reproducing}
K(\zeta_2,\zeta_1)=K^*(\zeta_1,\zeta_2)
=\int_{\Sigma_m}K(\zeta_2,\zeta)K(\zeta,\zeta_1)d\Sigma_m(\zeta)
=\frac{2\pi m^5}{\tilde{Z}_{l,m}}
\frac{K_1\bigl(m\sqrt{-(\zeta_2-\zeta_1^*)^2}\bigr)}
{m\sqrt{-(\zeta_2-\zeta_1^*)^2}},
\end{equation}
where $K_1(z)$ is the modified Bessel function of the first kind.
(Note that the difference between the normalisation factors appearing here and those in~\cite{Prugovecki:1995} is due to the fact that we are employing momentum variables exclusively, as opposed to velocity variables.)
Renormalising the kernel and taking the $l\rightarrow 0$ limit results (in the distributional sense) in the positive frequency Pauli--Jordan function~\cite[Appendix F]{Bogolubov+:1990} of the Klein--Gordon wave operator~\cite{Prugovecki:1996}
\begin{equation}
\lim_{l\rightarrow 0}\frac{i\tilde{Z}_{l,m}}{(2\pi)^3m^3}K(\zeta^\prime;\zeta)
\rightarrow K^{+}(q^\prime-q)
:=\frac{i}{(2\pi)^3}\int_{V_m^+}\exp\bigl(ik\cdot(q^\prime-q)\bigr)
d\Omega_m(k).
\end{equation}
Notice that for small arguments $K_2(x)\approx 2/x^2$, and therefore an infinite renormalisation has been performed in order to achieve this limit.

The alternative form of the inner product holding \textit{only} on elements of $\mathbf{P}L^2(\Sigma_m)$ (the normalisation constant is given by $\hat{Z}_{m}=m^2K_2(2lm)/K_1(2lm)$~\cite[page 105]{Prugovecki:1995})
\begin{subequations}
\begin{gather}
\langle \phi_1|\phi_2\rangle =i\frac{\hat{Z}_m}{m^2}
\int_{\Sigma_m}\phi_1^*(q,p)\overset{\leftrightarrow}{\partial}_{q^A}\phi_2(q,p)
d\sigma^A(q)d\Omega_m(p),\\
\phi_1^*(q,p)\overset{\leftrightarrow}{\partial}_{q^A}\phi_2(q,p):=
\phi_1^*(q,p)\partial_{q^A}[\phi_2(q,p)]
-\partial_{q^A}[\phi^*_1(q,p)]\phi_2(q,p),
\end{gather}
\end{subequations}
implies the existence of the system of covariance~\cite[Section 3.4]{Prugovecki:1992} made up of positive operator--valued measures
\begin{subequations}
\begin{equation}\label{eq:POV measures}
E(B)=i\frac{\hat{Z}_m}{m^2}\int_B
|\eta_{q,p}\rangle n^A\overset{\leftrightarrow}{\partial}_{q^A}
\langle\eta_{q,p}|\,d^4q\,d\Omega_m(p)
\end{equation}
($B\subset\Gamma$ is a Borel set), which satisfy the Poincar\'{e} covariance relations
\begin{equation}
U(\Lambda,b)E(B)U(\Lambda,b)^{-1}=E(\Lambda B+b).
\end{equation}
\end{subequations}
This in turn secures the interpretation of the integration of such measures over a Borel set $B\subset\Sigma_m$
\begin{equation}
\mathcal{P}_\phi(B)
=\langle\phi\vert E(B)\phi\rangle
=\int_{B\in\Sigma_m}\lvert(\mathbf{P}\phi)(q,p)\rvert^2d\Sigma_m(q,p),
\end{equation}
as the outcome of a simultaneous measurement of stochastic position and momentum.
The existence of the Poincar\'{e} covariant, conserved, probability current
\begin{subequations}
\begin{gather}
J^A(q):=\int_{V_m^+}\frac{2p^A}{m}\lvert\phi(q,p)\rvert^2 d\Omega_m(p),\\
U(\Lambda,b):J^A(q)\rightarrow (\Lambda J)^A\bigl(\Lambda^{-1}(q-b)\bigr),\quad
\partial_{q^A}J^A(q)=0,
\end{gather}
\end{subequations}
with positive--definite probability density $J^0$ related to the measures~\eqref{eq:POV measures}, not only secures the probabilistic interpretation of the theory, but also guarantees that the integration over a hypersurface in the inner product~\eqref{eq:inner product} can be taken over \textit{any} spatial hypersurface and not just spatial hyper-planes.
Since this will be of importance later on, it is useful to review the arguments leading to this result, which is a generalisation of that appearing in~\cite[pages 57-8]{Schweber:1961}.

First generalise the current by considering the product of any two elements of $\mathbf{P}L^2(\Sigma_m)$ 
\begin{equation}\label{eq:gen current}
J_{21}^A(q):=\int_{V_m^+}\frac{2p^A}{m}\phi^*_2(q,p)\phi_1(q,p) 
d\Omega_m(p),
\end{equation}
which due to the Lorentz invariance of the momentum measure, transforms as 
\begin{equation}\label{eq:J local}
J^A_{21}(q)\rightarrow 
\bigl(\Lambda(q)J_{21}\bigr)^A\bigl(\Lambda^{-1}(q)\bigl(q-b(q)\bigr)\bigr)
\end{equation}
under a local Poincar\'{e} transformation $\bigl(\Lambda(q),b(q)\bigr)$.
Using the momentum space representation and taking the partial derivative of~\eqref{eq:gen current} gives
\begin{equation}
\begin{split}
\partial_{q^A}J_{21}^A(q)=
\frac{2im^2}{\tilde{Z}_{l,m}}\int_{V_m^+} d\Omega_m(k_2)\int_{V_m^+} d\Omega_m(k_1)
\int_{V_m^+} d\Omega_m(p) \\
\tilde{\phi}_2^*(k_2)\tilde{\phi}_1(k_1)p\cdot(k_1-k_2)
\mathrm{e}^{-iq\cdot(k_2-k_1)}\mathrm{e}^{-\tfrac{l}{m}p\cdot(k_1+k_2)},
\end{split}
\end{equation}
which (once again using invariance of $d\Omega_m(p)$) vanishes when one boosts to the rest frame of $(k_1+k_2)$.
(Note that this probability current is conserved for any choice of resolution generator $\tilde{\eta}(k)$ provided that it is a real function of $k$~\cite[Section 2.8]{Prugovecki:1984}.)
Integrating this over a region $R\subset\mathbb{M}^4$ and making use of Gauss' law gives 
\begin{equation}\label{eq:surface integral}
\int_R d^4q\,\partial_{q^A}J^A_{21}(q)\equiv 0
=\oint_{\partial R} J_{21}(q)\cdot d\sigma(q),
\end{equation}
which, when specialised to the region between two hypersurfaces $\sigma^1$ and $\sigma^2$ (corresponding to $\Sigma_m^1$ and $\Sigma_m^2$ respectively) that coincide at spatial infinity, gives 
\begin{equation}
\int_{\Sigma_m^1} d\Sigma_m(q,p)\phi^*_1(q,p)\phi_2(q,p)
=\int_{\Sigma_m^2} d\Sigma_m(q,p)\phi^*_1(q,p)\phi_2(q,p).
\end{equation}

For later reference it is useful to consider the form of an integral over an arbitrary hypersurface $\sigma_{T_0}$ labelled by the constant value of a function of the $q$ variables $T(q)=T_0$.
The vector components of the normal to the surface is then defined by $n^A(q):=\eta^{AB}\partial_{q^B}[T(q)]/\eta^{CD}\partial_{q^C}[T(q)]\partial_{q^D}[T(q)]$, and the surface measure may be written as 
\begin{subequations}\label{eq:measures}
\begin{equation}
\int_{\Sigma_m}d\Sigma_m(q,p)\rightarrow
\int_{\mathbf{M}\times V_m^+}2p\cdot n(q)
\delta\bigl(T(q)-T_0\bigr)d^4q\,d\Omega_m(p).
\end{equation}
Introducing a new coordinate system $(T,\vec{x})$ in which the time variable is constant on $\sigma_{T_0}$, the gradient of $q$ with respect to the spatial variables $\partial_{\vec{x}}[q(x)]$ (the pullback of the embedding map~\cite{Kuchar:1976a}) gives the induced surface metric $\gamma_{ij}(x):=-\eta_{AB}\partial_{x^i}[q^A(x)]\partial_{x^j}[q^B(x)]$.
In this coordinate system the surface measure is given by
\begin{equation}
\int_{\sigma_{t_0}\times V_m^+}2p\cdot n(x)
\sqrt{\gamma(x)}\,d^3\vec{x}\,d\Omega_m(p)
=\int_{\sigma_{t_0}\times V_m^+}2p\cdot d\sigma(x)\,d\Omega_m(p),
\end{equation}
\end{subequations}
where the second form has been generalised to include the use of an arbitrary Lorentz frame; we find that the spatial measure has taken on the standard form~\cite[Appendix B]{Wald:1984}.

In the case where the $\phi$'s are replaced by the kernel~\eqref{eq:kernel} and we are therefore considering the reproducing property~\eqref{eq:reproducing}, we find a stronger result.
Considering any two spatial hypersurfaces $\Sigma_m^1$ and $\Sigma_m^2$ with a surface at spatial infinity $C$ with (outward--pointing) normal $\hat{r}$, the surface integral in~\eqref{eq:surface integral} then implies 
\begin{multline}\label{eq:repeater}
\int_{\Sigma_m^1} d\Sigma_m(q,p)K(\zeta_2;q,p)K(q,p;\zeta_1)\\
=\int_{\Sigma_m^2} d\Sigma_m(q,p)K(\zeta_2;q,p)K(q,p;\zeta_1)
+\int_{C\times V_m^+} 2p\cdot\hat{r}d\sigma(q)d\Omega_m(p)K(\zeta_2;q,p)K(q,p;\zeta_1).
\end{multline}
The fact that for large spacelike separations dominated by the spatial interval $\Delta\vec{q}$, $m\sqrt{-(\zeta_2-\zeta_1^*)^2}\approx m\lvert\Delta\vec{q}\rvert-il(p_2+p_1)\cdot\Delta\hat{q}$ and using the asymptotic behaviour of the modified Bessel function of the second kind~\cite[Equation 8.451 \#6, page 963]{Gradshteyn+Ryzhik:1980} $K_\nu(z)\approx \sqrt{\pi/(2z)}\mathrm{e}^{-z}$ as $z\rightarrow\infty$, we find that as the surface $C$ is taken to spatial infinity, it's contribution behaves as $\int dt\,d\Omega\,\mathrm{e}^{-2m\lvert\vec{q}\rvert}/(m^3\lvert\vec{q}\rvert)$, which vanishes in the limit.
Therefore for establishing the reproducing property of the kernel~\eqref{eq:kernel}, \textit{any} maximal spacelike hypersurface at all will suffice.
Note however that if points on the surface $C$ have light-like (or timelike) separation from points on both of $\Sigma_m^1$ and $\Sigma_m^2$, then we cannot conclude that its contribution vanishes in the limit since the kernel~\eqref{eq:reproducing} will no longer vanish as $C$ is taken to spatial infinity.

Before concluding this brief description of relativistic QG mechanics, we should make a few comments on interacting fields.
Introducing a relativistic potential (\textit{e.g.}, a vector potential in the case of the scattering of charged scalars off a background electro-magnetic field) involves the projection of the potential function onto $\mathbf{P}L^2(\Sigma_m)$, with the resulting dynamics given by an iterated kernel very similar to that which will appear when we consider a curved background spacetime~\eqref{eq:QGP}.
However, since the phenomena of pair creation in strong fields is an experimental fact, one introduces many--body, second--quantised QG field theory, in which the QG field operators are \textit{bona fide} densely defined operators on Fock space, in contrast to the conventional construction of local field theory.
Due to this fact, there are no divergences in the resulting perturbation series for the scattering matrix, and therefore no infinite renormalizations to be performed.
Furthermore, in the non-relativistic limit the role of the resolution generator $\eta$ in such measurements are interpretable as that of a confidence function, and the outcome of a measurement as a stochastic or average value only~\cite{Ali:1985}.
Then taking the sharp point limit $l\rightarrow 0$ leads identically to orthodox quantum mechanics~\cite[Section 2.6]{Prugovecki:1992}.
Taking the sharp point limit first however, leads to the formal perturbative series of local quantum field theory in any particular model, but \textit{only} after an infinite renormalisation is performed~\cite[Section 5.8]{Prugovecki:1995}.
Thus agreement of QG field theory models with the remarkably accurate predictions of the formally defined perturbative series of conventional local quantum field theory models~\cite{Donoghue+Golowich+Holstein:1992,Kinoshita:1990} is guaranteed at the formal level for small (but non--zero) choice of the fundamental scale $l$, which one expects on fundamental grounds should be of Planck length order~\cite{DeWitt:1962}.

\subsection{Geometrization}
\label{sect:geometrization}

The first step toward the generalisation of this structure to a curved spacetime manifold $\mathbf{M}$ is to note that one need not ever consider the underlying momentum space structure in~\eqref{eq:subirrep} at all, instead working completely within $\Gamma$~\cite[Section (2.2.7)]{Prugovecki:1984}.
Introducing the phase space wave function of the rest frame leads to the local wave functions
\begin{equation}
\eta_{m}(\zeta):=
\int_{V_m^+}\mathrm{e}^{-ik\cdot q}\tilde{\eta}(k\cdot p/m)
\tilde{\eta}(k) d\Omega_m(k)
=\int_{V_m^+}\tilde{\eta}^*_{\zeta}(k)\tilde{\eta}(k) d\Omega_m(k),
\end{equation}
and generating the wave functions of the boosted and translated frames through the representation of $\mathrm{ISO}_0(1,3)$ on the rest frame, leads to
\begin{equation}
\eta_{\zeta^\prime}(\zeta):=\bigl(U(\Lambda_{p^\prime/m},q^\prime)
\eta_{m}\bigr)(\zeta)
=K(\zeta;\zeta^\prime).
\end{equation}
Considering the action of an arbitrary group element on one of these frames, we find~\cite{Drechsler+Tuckey:1996}
\begin{subequations}\label{eq:group action}
\begin{equation}
\begin{split}
\bigl(U(\Lambda,b)\eta_{\zeta^\prime}\bigr)(\zeta)
=&\bigl(U(\Lambda,b)U(\Lambda_{p^\prime/m},q^\prime)\eta\bigr)(\zeta)\\
=&\bigl(U\bigl((\Lambda,b)
\star(\Lambda_{p^\prime/m},q^\prime)\bigr)\eta\bigr)(\zeta)
=\bigl(U(\Lambda\Lambda_{p^\prime/m},\Lambda q^\prime+b)\eta\bigr)(\zeta),
\end{split}
\end{equation}
and using the fact that $\Lambda\Lambda_{p/m}=\Lambda_{\Lambda p/m}\Lambda_R$ uniquely, where $\Lambda_R$ is the Wigner rotation and $\Lambda_{\Lambda p/m}$ is the boost to the velocity $\Lambda p/m$, we find that 
\begin{equation}
\begin{split}
\bigl(U(\Lambda,b)\eta_{q^\prime,p^\prime}\bigr)(\zeta)
=\bigl(U(\Lambda_{\Lambda p^\prime/m},\Lambda q^\prime+b)\eta\bigr)(\zeta)
=\eta_{(\Lambda,b)\star\zeta^\prime}(\zeta),
\end{split}
\end{equation}
\end{subequations}
where we have used the fact that the rest frame is rotationally invariant $U(\Lambda_R,0)\eta=\eta$.
Abstracting away the phase space functional dependence and considering $\eta_\zeta$ as a basis in $\mathbf{P}L^2(\Sigma_m)$, we define the set of quantum frames~\cite[Section~3.7]{Prugovecki:1992}
\begin{equation}\label{eq:quantum frames}
\begin{split}
\mathbf{Q}:=&\bigl\{\eta_\zeta\mid \zeta=q-i\tfrac{l}{m}p, 
q\in\mathbb{M}^4,p\in V_m^+\bigr\}\\
\equiv&\bigl\{\eta_\zeta\mid \eta_\zeta=\bigl(U(\Lambda_{p/m},q)\eta_m\bigr), 
q\in\mathbb{M}^4,p\in V_m^+\bigr\},
\end{split}
\end{equation}
with the group action from~\eqref{eq:group action}
\begin{equation}
U(g)\eta_{\zeta}=\eta_{g\star\zeta},\quad
g\in \mathrm{ISO}_0(1,3).
\end{equation}
The kernel~\eqref{eq:kernel} then plays the role of the overlap amplitude of these quantum frames
\begin{equation}\label{eq:overlap kernel}
\bigl\langle\eta_{\zeta^\prime}\big|\eta_\zeta\bigr\rangle=K(\zeta^\prime;\zeta)
\end{equation}
(and indeed may be considered as the local quantum metric in a curved spacetime~\cite[Section 5.2]{Prugovecki:1992}~\cite[Section 4.5]{Prugovecki:1995}).
The fact that $K(\zeta^\prime;\zeta)$ reproduces on $\mathbf{P}L^2(\Sigma_m)$ indicates that the operator defined by
\begin{equation}\label{eq:projection-identity}
\mathbf{P}_{\Sigma_m}:=\int_{\Sigma_m}
|\eta_\zeta\rangle d\Sigma_m(\zeta)\langle\eta_\zeta|,
\end{equation}
acts either as continuous resolution of the identity on $\mathbf{P}L^2(\Sigma_m)$, or a projection from $L^2(\Sigma_m)$ onto $\mathbf{P}L^2(\Sigma_m)$.
Therefore a scalar state may be decomposed on $\mathbf{Q}$ as
\begin{subequations}\label{eq:flat scalar}
\begin{equation}
\phi=\int_{\Sigma_m}d\Sigma_m(\zeta)\phi(\zeta)\eta_\zeta,
\end{equation}
where
\begin{equation}
\phi(\zeta):=
\langle\eta_{\zeta}|\phi\rangle
=\int_{\Sigma_m}\eta_\zeta^*(\zeta^\prime)
\phi(\zeta^\prime)d\Sigma_m(\zeta^\prime)
=\int_{\Sigma_m}K(\zeta;\zeta^\prime)
\phi(\zeta^\prime)d\Sigma_m(\zeta^\prime).
\end{equation}
\end{subequations}
(The structure described is actually a specific example of a more general procedure~\cite{Ali+Antoine+Gazeau:1993a,Ali+Antoine+Gazeau:1993b}.)

Since the frames in $\mathbf{Q}$ carry a representation of the Poincar\'{e} group, in order to relate the Hilbert space structure of $\mathbf{P}L^2(\Sigma_m)$ to a curved spacetime $\mathbf{M}$ the affine generalisation~\cite{Kobayashi+Nomizu:1963} $P\mathbf{M}$ of the Lorentz frame bundle $L\mathbf{M}$ over $\mathbf{M}$ is employed.
This bundle consists of all frames above each $x\in\mathbf{M}$ of the form $u=(e_A,\mathbf{a})$, where we will use a boldface letter to denote the vector itself $\mathbf{a}=a^Ae_A\in T\mathbf{M}$ (and continue to use, for example, $a$ to denote the vector components) and the linear frame is defined with respect to a particular coordinate frame via $e_A={E^\mu}_A\partial_\mu$.
The dual (coframe) basis is then given by $\theta^A=dx^\mu{E^A}_\mu$, where the vierbeins satisfy the following orthogonality relations ${E^A}_\mu{E^\mu}_B=\delta^A_B$, and ${E^\mu}_A{E^A}_\nu=\delta^\mu_\nu$.
The structure group of $P\mathbf{M}$ is the Poincar\'{e} group $\mathrm{ISO}(1,3)$ with right action on these frames defined by
\begin{equation}
u\star(\Lambda,b)=(e_A,\mathbf{a})\star(\Lambda,b)=(e^\prime_A,\mathbf{a}^\prime)
=\bigl(e_B{\Lambda^B}_A,(a^A+b^A)e_A\bigr).
\end{equation}
In practice we will actually employ $P_0\mathbf{M}$, constructed from Lorentz frames with future--pointing timelike and right--handed spatial basis vectors, which has $\mathrm{ISO}_0(1,3)$ as the structure group.

Implicit in this construction is the existence of a metric on $\mathbf{M}$ which has been used to reduce the affine frame bundle $A\mathbf{M}$ to $P\mathbf{M}$.
With this metric comes the unique, metric compatible, Levi--Civita connection, which is uniquely extendible to a connection on $P\mathbf{M}$~\cite[Section III.3]{Kobayashi+Nomizu:1963}.
In a given section (dependence on which will not be indicated) this connection determines the infinitesimal parallel transport, or covariant derivative, operator which acts on a tensor $T$ associated to $P\mathbf{M}$ through a tensor representation of the Poincar\'{e} group with generators ${M^A}_B$ and $P_A$ satisfying~\eqref{eq:generators}, along a path in $\mathbf{M}$ with tangent vector $X\in T\mathbf{M}$ as~\cite[Section 2.6]{Prugovecki:1992}
\begin{equation}
\nabla_XT=X^Ae_A[T]+\omega[X]T
=X^Ae_A[T]+\bigl(\tfrac{1}{2}{\omega^A}_B[X]{M^B}_A
+\hat{\theta}^A[X]P_A\bigr)T
,\quad X\in T\mathbf{M},
\end{equation}
where $\omega$ is the (Poincar\'{e}) Lie algebra--valued, Levi--Civita, connection one form acting in the representation of $T$, with action on a frame described by~\cite{Drechsler:1982}
\begin{subequations}\label{eq:nabla e}
\begin{gather}
\omega[X]:(e_A,\mathbf{a})\rightarrow
\bigl(X^B\Gamma^C_{BA}e_C,\bigl(X^A+\nabla_X[a]^A\bigr)e_A\bigr),\\
\nabla_X[a]^A:=X^B\bigl(e_B[a^A]+a^C\Gamma^A_{BC}\bigr).
\end{gather}
The components $\Gamma^A_{BC}$ appearing in~\eqref{eq:nabla e} are those of the torsion--free, Levi--Civita connection written in terms of the structure constants $[e_A,e_B]={C_{AB}}^Ce_C$ of the Lorentz frame $(e_A,0)$ as
\begin{gather}
\label{eq:Gammas}
\Gamma^A_{BC}={\omega^A}_C[e_B]=\tfrac{1}{2}({C_{BC}}^A+{C^A}_{BC}+{C^A}_{CB}),\\
{C_{BC}}^A:={E^A}_\mu\bigl(e_B[{E^\mu}_C]-e_C[{E^\mu}_B]\bigr),\quad
{C^A}_{BC}:=\eta^{AD}\eta_{CE}{C_{DB}}^E,
\end{gather}
\end{subequations}
and satisfy the metric compatibility condition $\omega_{(AB)}=0$.
The action of $\omega$ on the frame is an infinitesimal Poincar\'{e} transformation of the form
\begin{equation}
\omega[X]:(e_A,\mathbf{a})\rightarrow
(e_A,\mathbf{a})\star \bigl(\omega[X],X+\nabla_X[a]\bigr).
\end{equation}

Since parallel transport plays a fundamental role in QG field propagation in curved spacetimes, it is worthwhile at this point to review the computation of the results of parallel transport in a Poincar\'{e} frame bundle.
A frame that is parallel transported along a curve $\gamma$ with tangent $X\in T\mathbf{M}$ is one that does not change along $\gamma$, and may be written in a section $s=(e_A,\mathbf{a})$ as $(e_A,\mathbf{a})\star g_\gamma$.
Writing $\tau$ as a parameterisation of $\gamma$, we have $\partial_\tau[(e_A,a)\star g_\gamma]=0$ which gives~\cite[Section 10.1.4]{Nakahara:1990}
\begin{subequations}\label{eq:pt equation}
\begin{equation}
(e_A,\mathbf{a})\star \bigl(\omega[X],X+\nabla_X[a]\bigr)\star g_\gamma
+(e_A,\mathbf{a})\star X[g_\gamma]=0,
\end{equation}
and so the transition element satisfies
\begin{equation}
\partial_\tau[g_\gamma]=-\bigl(\omega[X],X+\nabla_X[a]\bigr)\star g_\gamma.
\end{equation}
\end{subequations}
It is convenient to transfer this action to the components of a vector in the Mobius representation~\cite{Hehl+:1995} where $V\in T\mathbf{M}$ is written as a column vector in $\mathbb{R}^5$ and the infinitesimal Poincar\'{e} transformation as a $5\times 5$ matrix, which acts on $V$ as
\begin{equation}\label{eq:Mobius}
(\kappa,\beta)\star X\rightarrow
\begin{bmatrix}\kappa&\beta\\0&0\end{bmatrix}
\begin{bmatrix}V\\1\end{bmatrix},\quad
\kappa=-\omega[X],\quad 
\beta=-\bigl(X+\nabla_X[a]\bigr).
\end{equation}
The restriction $\omega_{(AB)}=0$ ensures that this may be looked upon as an infinitesimal Poincar\'{e} transformation with generators that satisfy the Lie algebra of the Poincar\'{e} group.

In this representation the exponential map (of $\mathfrak{iso}(1,3)$ into $\mathrm{ISO}(1,3)$) may be split up into a Lorentz boost and a translation by a straightforward induction argument
\begin{equation}
\exp(\kappa,\beta)=\sum^\infty_{n=0}\frac{1}{n!}(\kappa,\beta)^n
=\bigl(\exp(\kappa),(\exp(\kappa)-1)\kappa^{-1}\beta\bigr),
\end{equation}
where $\exp(\kappa)$ is the exponential map of $\mathfrak{so}(1,3)$ to $\mathrm{SO}(1,3)$.
Also by induction, the product of $N$ such exponential maps may be re--represented as
\begin{equation}
\prod_{k=N}^1\exp(\kappa_k,\beta_k)
=\Biggl(\prod_{k=N}^1\exp(\kappa_k),
\sum_{m=1}^N\biggl(\prod_{k=N}^{m+1}\exp(\kappa_k)\biggr)
\bigl(\exp(\kappa_m)-1\bigr)\kappa_m^{-1}\beta_m\Biggr).
\end{equation}
Note that the product is ordered: $\prod_{n=N}^1g_n:=g_N\cdot g_{N-1}\cdots g_2\cdot g_1$, and that $(\exp(\kappa)-1)\kappa^{-1}:=\bigl(\sum_{n=0}^\infty\kappa^n/(n+1)!\bigr)$.

If we consider parallel transport along a curve $\gamma:[\tau_i,\tau_f]\subset\mathbb{R}\rightarrow\mathbf{M}$, one finds~\cite{Nakahara:1990} that the finite $\mathrm{ISO}_0(1,3)$ parallel transport map that is generated by the infinitesimal connection may be written as
\begin{equation}
g\bigl(\gamma(\tau_f,\tau_i)\bigr)
=\mathrm{P}\exp\Bigl(\int_{\tau_i}^{\tau_f}\bigl(\kappa(\tau),\beta(\tau)\bigr)d\tau\Bigr),
\end{equation}
where $\mathrm{P}$ is the path ordering operator (larger values of the parameter $\tau$ appearing to the left).
Computing this `path integral' by splitting up the time interval up into N equal pieces~\cite{Prugovecki:1995}: $\Delta \tau=(\tau_f-\tau_i)/N=\tau_n-\tau_{n-1}$, $\tau_i=\tau_0$, $\tau_f=\tau_N$, $\tau_n=\tau_i+n\Delta \tau$ and writing
\begin{equation}
g_\gamma(\tau_{n+1},\tau_n)
=\mathrm{P}\exp\Bigl(\int_{\tau_{n-1}}^{\tau_n}\bigl(\kappa(\tau),\beta(\tau)\bigr)d\tau\Bigr)
\approx \exp\Bigl(\bigl(\kappa(\tau_n),\beta(\tau_n)\bigr)\Delta \tau\Bigr),
\end{equation}
noting that the evolution along $\gamma$ is composable: $g_\gamma(\tau_n,\tau_{n-2})=g_\gamma(\tau_n,\tau_{n-1})g_\gamma(\tau_{n-1},\tau_{n-2})$, and taking the $N\rightarrow\infty$ limit in which the approximation becomes exact, the transition element is given by
\begin{equation}\label{eq:transition element}
g_\gamma(\tau_N,\tau_0)=\lim_{N\rightarrow\infty}\prod_{n=N-1}^0
\exp\Bigl(\bigl(\kappa(\tau_n),\beta(\tau_n)\bigr)\Delta \tau\Bigr).
\end{equation}
Using the above results for the representation of the Lie algebra, this is equal to (in the $N\rightarrow\infty$ limit)
\begin{equation}\label{eq:geodesic exponential}
g_\gamma(\tau_N,\tau_0)=
\Bigl(\mathrm{P}\exp\Bigl(
\int_{\tau_i}^{\tau_f}\kappa(\tau)d\tau\Bigr),
\int_{\tau_i}^{\tau_f}d\tau\,\mathrm{P}\exp\Bigl(
\int_{\tau}^{\tau_f}\kappa(\tau^\prime)d\tau^\prime\Bigr)\beta(\tau)\Bigr).
\end{equation}
In practice these results will be applied to geodesics that are forward--pointing in the chosen time coordinate, so that the path ordering may be replaced by time ordering; however in general note that the $P$ operator disappears in cases where $[\kappa(\tau),\kappa(\tau^\prime)]=0$.

It will be useful to display the behaviour of this transition element under a change of Poincar\'{e} frame.
Writing the transition element corresponding to parallel transport along a path beginning at $x$ and ending at $x^\prime$ in a section $s^\prime$ as $g^{s^\prime}_{\gamma}(x^\prime,x)$, if the section is related to another choice by $s^\prime(x)=s(x)\star g(x)$, from the structure of~\eqref{eq:pt equation} we see that the transition elements are related by~\cite{Nakahara:1990}
\begin{subequations}\label{eq:transition relation}
\begin{equation}
g^{s}_{\gamma}(x^\prime,x)
=g(x^\prime)\star
g^{s^\prime}_{\gamma}(x^\prime,x)\star g(x)^{-1}.
\end{equation}
In particular, the relationship between an arbitrary Poincar\'{e} section $s=\bigl(e_A,\mathbf{a}\bigr)$ and the related Lorentz section $s_0:=(e_A,0)$ is
\begin{equation}
g^{s}_{\gamma}(x^\prime,x)
=\bigl(\mathbf{1},\mathbf{a}(x^\prime)\bigr)^{-1}\star
g^{s_0}_{\gamma}(x^\prime,x)\star
\bigl(\mathbf{1},\mathbf{a}(x)\bigr),
\end{equation}
\end{subequations}
where we have used $s=s_0\star(\mathbf{1},\mathbf{a})$.

\subsection{Quantum Propagation}
\label{sect:QP}

With the Poincar\'{e} frame bundle $P_0\mathbf{M}$ and the set of quantum frames $\mathbf{Q}$~\eqref{eq:quantum frames}, we have what may be considered as the classical and quantum counterparts of the same physical entity, namely, the frames of reference on which properties of the physical system are measured.
Since both sets of frames transform under a representation of the Poincar\'{e} group, the problem of how to relate the two sets is resolved by defining the principle quantum frame bundle over $\mathbf{M}$ as $Q_0\mathbf{M}=P_0\mathbf{M}\times_{\mathrm{ISO}_0(1,3)}\mathbf{Q}$, which is the G--product~\cite[page 54]{Kobayashi+Nomizu:1963} of the Poincar\'{e} frame bundle with $\mathbf{Q}$.
A typical element of $Q_0\mathbf{M}$ is given by the associated quantum frame
\begin{equation}
\Phi^{u(x)}_\zeta=\bigl(u(x),\eta_\zeta\bigr),
\end{equation}
which is a pair consisting of a Poincar\'{e} frame $u(x)$ and a quantum frame $\eta_\zeta$.
The fibre bundle structure of $Q_0\mathbf{M}$ is completed by defining the projection map back to the underlying manifold
\begin{equation}
\pi\bigl(\Phi^{u(x)}_\zeta\bigr)=x\in\mathbf{M},
\end{equation}
and the local trivialisation map onto the typical fibre which is isomorphic to $\mathbf{P}L^2(\Sigma_m)$
\begin{equation}
\sigma^{u(x)}\bigl(\Phi^{u(x)}_\zeta\bigr)
=\sigma^{u(x)}\bigl(u(x),\eta_\zeta\bigr)=\eta_\zeta.
\end{equation}
Equivalence classes are identified via the left action
\begin{subequations}\label{eq:equiv}
\begin{equation}
g\star(u,\eta_\zeta)
=\bigl(u\star g^{-1},U(g)\star\eta_\zeta\bigr)
=\bigl(u\star g^{-1},\eta_{g\star\zeta}\bigr),
\end{equation}
from which it is clear that 
\begin{equation}
\sigma^{u\star g^{-1}}=U(g)\sigma^u,\quad
\bigl(\sigma^{u\star g^{-1}}\bigr)^{-1}=(\sigma^u)^{-1}U(g^{-1}).
\end{equation}
\end{subequations}

This product `solders' the quantum frames $\mathbf{Q}$ to the Poincar\'{e} frames $P_0\mathbf{M}$ of $\mathbf{M}$, and parallel transport as defined by the Levi-Civita connection in $P_0\mathbf{M}$ is then transferable to the frames in $\mathbf{Q}$.
Given a section $s(x)=\bigl(e_A(x),\mathbf{a}(x)\bigr)$ of $P_0\mathbf{M}$, the associated quantum frames are given by
\begin{equation}
\Phi^{s(x)}_\zeta=\bigl(\sigma^{s(x)}\bigr)^{-1}\eta_\zeta,
\end{equation}
which on account of~\eqref{eq:equiv} satisfies
\begin{equation}
\Phi^{s\star g}_\zeta=(\sigma^{s\star g})^{-1}\eta_\zeta
=(\sigma^s)^{-1}U(g)\eta_\zeta
=(\sigma^s)^{-1}\eta_{g\star\zeta}
=\Phi^{s}_{g\star \zeta}.
\end{equation}
For simplicity we will adopt the notation
\begin{equation}\label{eq:local kernel}
\Bigl\langle\Phi^{s(x)}_{\zeta^\prime}\Big|\Phi^{s(x)}_\zeta\Bigr\rangle
:=\Bigl\langle\sigma^{s(x)}\Phi^{s(x)}_{\zeta^\prime}\Big|
\sigma^{s(x)}\Phi^{s(x)}_\zeta\Bigr\rangle
=\langle\eta_{\zeta^\prime}|\eta_\zeta\rangle
=K(\zeta^\prime;\zeta)
\end{equation}
for the overlap amplitude~\eqref{eq:overlap kernel} of quantum frames in a given fibre of $Q\mathbf{M}$.
Similarly, the continuous resolution of the identity~\eqref{eq:projection-identity} becomes an identity in each fibre of $Q_0\mathbf{M}$
\begin{equation}\label{eq:local identity}
\mathbf{P}^{s(x)}_{\Sigma_m}=\int_{\Sigma_m}
\Big|\Phi^{s(x)}_\zeta\Bigr\rangle d\Sigma_m(\zeta)
\Bigl\langle\Phi^{s(x)}_\zeta\Big|.
\end{equation}

Then, if the outcome of parallel transport along a path $\gamma$ connecting two points $x,x^\prime\in\mathbf{M}$ in the section $s(x)$ as computed in section~\ref{sect:geometrization} is the map $s(x)\rightarrow s(x^\prime)\star g_\gamma(x^\prime,x)$, then this induces the change of quantum frame
\begin{equation}
\Phi^{s(x)}_\zeta\rightarrow
\tau_\gamma(x^\prime,x)\Phi^{s(x)}_\zeta
=\Phi^{s(x^\prime)\star g_\gamma(x^\prime,x)}_\zeta
=\Phi^{s(x^\prime)}_{g_\gamma(x^\prime,x)\star\zeta},
\end{equation}
from which we define the propagator for parallel transport along $\gamma$ in the section $s$ as the amplitude
\begin{equation}\label{eq:K general}
K_{\gamma(x^\prime,x)}^s(x^\prime,\zeta^\prime;x,\zeta):=
\Bigl\langle \Phi^{s(x^\prime)}_{\zeta^\prime}\Big|
\tau_\gamma(x^\prime,x)\Phi^{s(x)}_\zeta\Bigr\rangle 
=\Bigl\langle \Phi^{s(x^\prime)}_{\zeta^\prime}\Big|
\Phi^{s(x^\prime)}_{g_\gamma(x^\prime,x)\star\zeta}\Bigr\rangle 
=K\bigl(\zeta^\prime;g_\gamma(x^\prime,x)\star\zeta\bigr),
\end{equation}
and from~\eqref{eq:kernel} we also know that $K(\zeta^\prime,g\star\zeta)=K(g^{-1}\star\zeta^\prime,\zeta)$.
An important consequence of~\eqref{eq:K general} is the behaviour of the propagator under changes of Poincar\'{e} gauge.
From 
\begin{subequations}
\begin{equation}\label{eq:K gauge}
K_{\gamma(x^\prime,x)}^{s\star g}(x^\prime,\zeta^\prime;x,\zeta)=
K_{\gamma(x^\prime,x)}^{s}\bigl(x^\prime,g(x^\prime)\star\zeta^\prime;
x,g(x)\star\zeta\bigr),
\end{equation}
we deduce that 
\begin{equation}\label{eq:consistency}
\begin{split}
K\bigl(\zeta^\prime;g^{s\star g}_\gamma(x^\prime,x)\star\zeta\bigr)
=&K\bigl(g(x^\prime)\star\zeta^\prime;
g^{s}_\gamma(x^\prime,x)\star g(x)\star\zeta\bigr)\\
=&K\bigl(\zeta^\prime;
g(x^\prime)^{-1}\star g^{s}_\gamma(x^\prime,x)\star g(x)\star\zeta\bigr),
\end{split}
\end{equation}
\end{subequations}
where $g_\gamma^s$ is the transition element from parallel transport in the Poincar\'{e} section $s$.
Clearly~\eqref{eq:consistency} expresses the consistency of the action of the Poincar\'{e} group on the Poincar\'{e} frames~\eqref{eq:transition relation} and the propagator.
An important special case of this that will arise later is when we take $s_L=(e_A,0)$ and $g=\bigl(\mathbf{1},a(x)\bigr)$, and we are therefore relating the amplitude in a Poincar\'{e} gauge to the related Lorentz frame.

Notice that the construction of $Q_0\mathbf{M}$ has attached the entire Hilbert space $\mathbf{P}L^2(\Sigma_m)$ to each point $x\in\mathbf{M}$, and the reproducing kernel~\eqref{eq:local kernel} and therefore the resolution of the identity~\eqref{eq:local identity} act within each fibre of $Q_0\mathbf{M}$.
Although the entire set of states in each fibre is not easily physically interpretable (unless spacetime is flat, in which case each fibre may be identified~\cite{Prugovecki:1996} as will be discussed further below), since the formalism depends on parallel transport and the local overlap amplitudes~\eqref{eq:K general} it is necessary to have the entire space reproduced locally.
The states that \textit{are} interpretable in a straightforward manner are those which are stochastically localised at the point of contact of the Poincar\'{e} frame with $\mathbf{M}$, and are therefore described in a section $s=\bigl(e_A,\mathbf{a}(x)\bigr)$ by the collection~\cite{Prugovecki:1994,Prugovecki:1995,Prugovecki:1996}
\begin{equation}\label{eq:q section}
\hat{\zeta}(x):=-a(x)-i\tfrac{l}{m}p,\quad
p\in V_m^+,
\end{equation}
where recall that a non--zero $\mathbf{a}(x)$ indicates that the section is a translation of the related Lorentz section $s_L(x):=(e_A,0)$ by $-a(x)$, and so transforming~\eqref{eq:q section} back to the related Lorentz frame results in the collection of quantum frames of all possible momenta stochastically localised at the origin of the frame $\hat{\zeta}(x)=-i\tfrac{l}{m}p$.

The assignment~\eqref{eq:q section} is unique in any section of $P_0\mathbf{M}$, and given a change of frame 
\begin{subequations}\label{eq:change of frame}
\begin{equation}
s(x)=\bigl(e_A(x),\mathbf{a}(x)\bigr)
=\bigl(e_A^\prime(x),\mathbf{a}^\prime(x)\bigr)
\star\bigl(\Lambda(x),b(x)\bigr)
=:s^\prime(x)\star\bigl(\Lambda(x),b(x)\bigr),
\end{equation}
the action is transferred to the quantum frame
\begin{equation}
\hat{\zeta}(x)\rightarrow
\bigl(\Lambda(x),b(x)\bigr)\star\hat{\zeta}(x)
=\Lambda(x)(-a(x)-i\tfrac{l}{m}p)+b(x)
=-a^\prime(x)-i\tfrac{l}{m}p^\prime,
\end{equation}
\end{subequations}
where we have identified the components of the momentum in the primed frame $p^\prime=\Lambda(x)p$ as well as the components of the translation vector $a^\prime(x)=\Lambda(x)a(x)-b(x)$.
This recognises the fact that due to the nature of the construction of the bundle $Q_0\mathbf{M}$, in general $\zeta$ may be considered as a vector in the complexified tangent space $T\mathbf{M}^{\mathbb{C}}$
\begin{equation}
\pmb{\zeta}:=\bigl(q-a(x)-i\tfrac{l}{m}p\bigr)^A e_A(x)
=\bigl(q-i\tfrac{l}{m}p\bigr)^A e_A(x)-\mathbf{a}(x).
\end{equation}
The transformation property~\eqref{eq:change of frame} implies that the form of the propagator for parallel transport in a section $s$ is invariant under an arbitrary change of frame
\begin{equation}\label{eq:gauge invariance}
K_{\gamma(x^\prime,x)}^{s}\bigl(x^\prime,\hat{\zeta}(x^\prime)
;x,\hat{\zeta}(x)\bigr)
=K_{\gamma(x^\prime,x)}^{s\star g^{-1}}
\bigl(x^\prime,g(x^\prime)\star\hat{\zeta}(x^\prime)
;x,g(x)\star\hat{\zeta}(x)\bigr),
\end{equation}
and guarantees the consistency of the choice of local frame with poincar\'{e} invariance, where the frame is transported as before, and the outcome of parallel transport along a path $\gamma$ between frames localised at $x^\prime$ and $x$ is defined by
\begin{equation}\label{eq:gauge propagator}
\begin{split}
K_{\gamma(x^\prime,x)}^{s}\bigl(x^\prime,\hat{\zeta}(x^\prime);
x,\hat{\zeta}(x)\bigr):=&
\Bigl\langle \Phi^{s(x^\prime)}_{\hat{\zeta}(x^\prime)}\Big|
\tau_\gamma(x^\prime,x)\Phi^{s(x)}_{\hat{\zeta}(x)}\Bigr\rangle 
=\Bigl\langle \Phi^{s(x^\prime)}_{\hat{\zeta}(x^\prime)}\Big|
\Phi^{s(x^\prime)}_{g_\gamma(x^\prime,x)
\star\hat{\zeta}(x)}\Bigr\rangle \\
=&K\bigl(\hat{\zeta}(x^\prime);
g_\gamma(x^\prime,x)\star\hat{\zeta}(x)\bigr).
\end{split}
\end{equation}

The picture that emerges from this is that the Poincar\'{e} frame $(e_A(x),\mathbf{a}(x))$ related to the classical spacetime manifold is replaced by the set of quantum states $\hat{\zeta}(x)$ which represents the set of states of all possible momenta which are stochastically localised at $x$.
Combined with the exponential map that generates Riemann normal coordinates in a suitably small neighbourhood of a point (limited by the magnitude of the local curvature~\cite[Sections 8.6 and 11.6]{MTW:1973}), this also allows an interpretation of the quantum frame wave function $\eta_{\zeta^\prime}(\zeta)$ which is expressed in terms of flat, tangent space coordinates, as the wave function of an extended state in spacetime in accordance with the discussion in Section~\ref{sect:RQM}.
It is in this sense that QG propagation is said to be geometrically local (\textit{i.e.}, occurring within individual fibres \textit{above} $\mathbf{M}$) but not topologically local (\textit{i.e.}, not interpretable as arbitrarily precisely localizable quantum states \textit{in} $\mathbf{M}$).

Although the propagator for parallel transport~\eqref{eq:K general} or~\eqref{eq:gauge propagator} provides an appropriate description of quantum propagation in the case of Minkowski spacetime (and, as we will see, in any frame of reference whatsoever), due to the path dependence of parallel propagation in curved spacetimes, in general it does not provide a reproducing kernel.
The rectification of this situation is to construct the \textit{quantum--geometric propagator}~\cite[Section 4.6]{Prugovecki:1995} by iterating the propagator~\eqref{eq:K general} over all broken, forward--pointing, geodesic paths between the initial and final points, in analogy with the computation of the transition element in Section~\ref{sect:geometrization}.

Begin by considering a foliation of the spacetime into spacelike hypersurfaces $\sigma_t$ labelled by the value of a time parameter $t$.
Given an initial point $x_i\in\sigma_{t_i}$ and a final point $x_f\in\sigma_{t_f}$, a sequence of hypersurfaces $\sigma_{t_n}$ is defined by the finite time intervals as before: $\Delta t=(t_f-t_i)/N=t_n-t_{n-1}$, $t_0:=t_i$, $t_N:=t_f$, and using the propagators for parallel transport along geodesics $\gamma(x_n,x_{n-1})$ with endpoints $x_n\in\sigma_{t_n}$, the quantum--geometric propagator is defined as the limit~\cite{Prugovecki:1996}
\begin{subequations}\label{eq:QGP}
\begin{multline}
\mathbf{K}^s\bigl(x_f,\hat{\zeta}(x_f);x_i,\hat{\zeta}(x_i)\bigr):=\\
\lim_{N\rightarrow\infty}
\prod^1_{n=N-1}
\int_{\sigma_{t_n}\times V_m^+}d\Sigma_m(x_n,p_n)
K^s_{\gamma(x_{n+1},x_n)}\bigl(x_{n+1},\hat{\zeta}(x_{n+1});
x_n,\hat{\zeta}(x_n)\bigr)\\
\times K^s_{\gamma(x_1,x_0)}\bigl(x_1,\hat{\zeta}(x_1);
x_0,\hat{\zeta}(x_0)\bigr),
\end{multline}
where by construction, this satisfies the curved spacetime analogue of~\eqref{eq:reproducing} (where $t_2>t>t_1$)
\begin{equation}
\begin{split}
\mathbf{K}^s\bigl(x_2,\hat{\zeta}(x_2);x_1,\hat{\zeta}(x_1)\bigr)
=&{\mathbf{K}^s}^*\bigl(x_1,\hat{\zeta}(x_1);x_2,\hat{\zeta}(x_2)\bigr)\\
=&\int d\Sigma(x,p)
\mathbf{K}^s\bigl(x_2,\hat{\zeta}(x_2);x,\hat{\zeta}(x)\bigr)
\mathbf{K}^s\bigl(x,\hat{\zeta}(x);x_1,\hat{\zeta}(x_1)\bigr).
\end{split}
\end{equation}
\end{subequations}
The measure appearing in equation~\eqref{eq:QGP} is defined by $d\Sigma_m(x,p):=2p\cdot d\sigma(x)d\Omega_m(p)$, and although at first sight the resulting construction would appear depend on the gauge to a greater degree than does~\eqref{eq:gauge invariance}, the fact that it is not may be seen by making use of~\eqref{eq:change of frame}.
The reproducing property of the QG--propagator embodies the spirit of the `principle of path independence' as formulated by Teitelboim~\cite{Teitelboim:1973,Kuchar:1988} and extended by the present author~\cite{Clayton:1996b,Clayton:1997c}, although the extent to which results depend on the choice of foliation is not clear at this time.

Even though we have been primarily describing parallel transport and QG propagation in $Q_0\mathbf{M}$, it is easily adapted to a scalar field by constructing the Klein--Gordon quantum bundle $E_0\mathbf{M}=Q_0\mathbf{M}\times_{\mathrm{ISO}_0(1,3)}\mathbf{P}L^2(\Sigma_m)$.
A section of $E_0\mathbf{M}$ would represent an assignment of scalar field modes above each point in $\mathbf{M}$ expanded in terms of the quantum frames as
\begin{equation}
\pmb{\phi}_x=\int_{\Sigma_m}\phi_x(\zeta)\Phi^{s(x)}_\zeta d\Sigma_m(\zeta),
\end{equation}
which is just~\eqref{eq:flat scalar} expressed in each fibre of $Q_0\mathbf{M}$.
Alternatively, making use of the identification of the stochastically localised frames, a complete description of initial data for a scalar field is given by
\begin{equation}\label{eq:Cauchy data}
\pmb{\phi}_{\sigma}=
\int_{\sigma\times V_m^+}\phi\bigl(x,\hat{\zeta}(x)\bigr)
\Phi^{s(x)}_{\hat{\zeta}(x)} d\Sigma_m(x,p).
\end{equation}
Parallel transport is transferred to the local components by considering the amplitudes to find the scalar field in those modes that are stochastically localised about the final point 
\begin{equation}
\begin{split}
\pmb{\tau}_\gamma(x,x_0):\pmb{\phi}_{\sigma_{t_0}}\rightarrow
\pmb{\phi}^\parallel_{\sigma_{t}}
=&\int_{\sigma_{t_0}\times V_m^+}\phi\bigl(x_0,\hat{\zeta}(x_0)\bigr)
\tau_\gamma(x,x_0)\Phi^{s(x_0)}_{\hat{\zeta}(x_0)} d\Sigma_m(x_0,p)\\
\rightarrow&\int_{\sigma_{t_0}\times V_m^+}\phi\bigl(x_0,\hat{\zeta}(x_0)\bigr)
K_\gamma^{s(x)}\bigl(x,\hat{\zeta}(x);x_0,\hat{\zeta}(x_0)\bigr)
\Phi^{s(x)}_{\hat{\zeta}(x)} d\Sigma_m(x_0,p), 
\end{split}
\end{equation}
where in the second form the result of parallel transport has been projected back onto the stochastically localised modes.
The iteration of this parallel transport operation over all forward--pointing geodesic arcs reproduces~\eqref{eq:QGP}, and the time evolution of the initial data is given by replacing the propagator for parallel transport with the quantum--geometric propagator
\begin{equation}
\pmb{\phi}_{\sigma_{t}}
=\int_{\sigma_{t_0}\times V_m^+}\phi\bigl(x_0,\hat{\zeta}(x_0)\bigr)
\mathbf{K}^{s}\bigl(x,\hat{\zeta}(x);x_0,\hat{\zeta}(x_0)\bigr)
\Phi^{s(x)}_{\hat{\zeta}(x)} d\Sigma_m(x_0,p),
\end{equation}
generating a section of $E_0\mathbf{M}$ from the initial data~\eqref{eq:Cauchy data}.

It is worthwhile at this stage to make a few comments on the integrals appearing in~\eqref{eq:QGP}.
In Minkowski spacetime, the reproducing property~\eqref{eq:repeater} may be rewritten in the following way
\begin{equation}\label{eq:gfy}
\int_{\Sigma_m}d\Sigma_m(\zeta)K(\zeta^{\prime\prime},\zeta)
K(\zeta,\zeta^\prime)
=\int_{\sigma\times V_m^+} d\Sigma_m(x,p)
K\bigl(\zeta^{\prime\prime},g_\gamma(x,x_0)\star\hat{\zeta}(x_0)\bigr)
K\bigl(g_\gamma(x,x_0)\star\hat{\zeta}(x_0),\zeta^\prime\bigr),
\end{equation}
where we have used the fact that the outcome of parallel transport in this case just gives the separation vector between the two points (this will be discussed in more detail below), and so $x_0$ may be chosen arbitrarily as all dependence on it will cancel identically.
In~\eqref{eq:gfy} we recognise the form of the surface integral in~\eqref{eq:measures}, and in fact have a special case of the integrals appearing in~\eqref{eq:QGP}.
Thus we find that what is happening (operationally) in~\eqref{eq:QGP} is that the outcome of parallel transport $g_\gamma(x,x_0)\star\hat{\zeta}(x_0)$ is (at least for spacetimes with small enough curvature) a map from the hypersurface $\sigma$ that spans $\mathbb{R}^3$, and may therefore be considered to be a surface integral in the same sense as~\eqref{eq:gfy}.
(Note that the outcome of the integration in curved spacetimes will depend in general on $x_0$ in a nontrivial way, so the kernel is not expected to reproduce identically.
Physically the iteration may be viewed in analogy to a multi--slit experiment: a quantum state, initially stochastically localised on the Cauchy surface $\sigma_{t_i}$ will propagate indiscriminantly along all possible broken geodesic paths to a final state localised on $\sigma_{t_f}$.

It is useful to see how this construction proceeds over a flat (Minkowski) spacetime, however considering an arbitrary coordinate chart (assumed for simplicity to be global), frame of reference and family of hypersurfaces on which to construct the quantum geometric propagator.
To begin, note~\cite{Drechsler+Tuckey:1996} that any section of the Poincar\'{e} frame bundle may be written as an arbitrary Poincar\'{e} transformation of the parallel propagation of a Lorentz frame at $O$ (taken as the origin of coordinates) as $s(x):=s_O(x)\star g_O(x)$, where $s_O(x):=(e_A^O(x),0)$ and $e_A^O(x)$ is the chosen frame at $O$ parallelly transported to all points of $\mathbf{M}$, expressed in terms of some arbitrary coordinate system. 
(This can easily be generalised to the parallelly propagated frame $s_\parallel(x):=\bigl(e_A^O(x),-\mathbf{X}_M(x)\bigr)$, where $\mathbf{X}_M$ is the Minkowski coordinate vector of the point $x\in\mathbf{M}$ again expressed in some arbitrary coordinates).
If the arbitrary section is written as $s(x)=\bigl(e^\prime_A(x),\mathbf{a}(x)\bigr)$, then the choice of Poincar\'{e} transformation may be written as $g_O(x)=\bigl(\Lambda_O(x),0\bigr)\star\bigl(\mathbf{1},\mathbf{a}(x)\bigr)$, and so~\eqref{eq:K gauge} relates the propagator in $s$ to that in $s_O$
\begin{subequations}
\begin{equation}
\begin{split}
K^s_{\gamma}\bigl(x^\prime,\hat{\zeta}(x^\prime);x,\hat{\zeta}(x)\bigr)
=&K\bigl(\hat{\zeta}(x^\prime);g^s_\gamma(x^\prime,x)\star\hat{\zeta}(x)\bigr)\\
=&K\bigl(\hat{\zeta}(x^\prime);
g_O(x^\prime)^{-1}\star g^{s_O}_\gamma(x^\prime,x)\star g_O(x)
\star\hat{\zeta}(x)\bigr)\\
=&K\bigl(-i\tfrac{l}{m}\Lambda_O(x^\prime)p^\prime;
g^{s_O}_\gamma(x^\prime,x)\star(-i\tfrac{l}{m}\Lambda_O(x)p)\bigr),
\end{split}
\end{equation}
where we have used the fact that by definition $\hat{\zeta}(x)=\bigl(\mathbf{1},\mathbf{a}(x)\bigr)^{-1}\star(-i\tfrac{l}{m}p)$.
Since we know that the outcome of parallel transport in $s_O$ is given by $g^{s_O}(x^\prime,x)=\bigl(\mathbf{1},X_M(x)-X_M(x^\prime)\bigr)$ where $X_M(x)$ is coordinate components of the point $x$ in the Minkowski coordinate system related to $O$, the propagator in a generic frame $s$ may be written as
\begin{gather}\label{eq:flat propagator}
\mathbf{K}^s\bigl(x_2,\hat{\zeta}_2(x_2);x_1,\hat{\zeta}_1(x_1)\bigr)
=K^s_{\gamma}\bigl(x_2,\hat{\zeta}_2(x_2);x_1,\hat{\zeta}_1(x_1)\bigr)
=K\bigl(\tilde{\zeta}_2(x_2);\tilde{\zeta}_1(x_1)\bigr),\\
\tilde{\zeta}_n(x_n):=X_M(x_n)-i\tfrac{l}{m}\Lambda_O(x_n)p_n.
\end{gather}
\end{subequations}

This result has some very intriguing features; not only is the spatial dependence identical to that of the stochastic quantum mechanics result~\eqref{eq:kernel} (and so equivalent to the hyper-plane results quoted in~\cite{Prugovecki:1996,Coleman:1996,Drechsler+Tuckey:1996}), but the only difference between the results in a general frame and those in a hyper-plane slicing is the presence of the Lorentz factors acting on the momenta at the endpoints.
This is not surprising since one should be able to view the propagator in a flat spacetime as a boost back to the Minkowski frame, a trivial parallel transport that just induces a translation factor, and a boost back to the general frame at the endpoint of propagation.
Indeed, right away we see that what replaces the Bogolubov transformations of the conventional local quantum field theory on a curved background~\cite{DeWitt:1975}, is a physically reasonable shift of the momentum spectrum of the state.
As a specific example, consider an initial state that is at rest at $O$ in the Minkowski frame.
The amplitude to find the state stochastically localised about a point $x$ with stochastic momentum $p$ on a later hypersurface $\sigma_t$ is evidently given by
\begin{equation}
\phi\bigl(x,\hat{\zeta}(x)\bigr)=K\bigl(\tilde{\zeta}(x);-iln^0\bigr),
\end{equation}
where $n^0=(1,\vec{0})$ and $x\in\sigma_t$.
Clearly only the momentum spectrum is distorted by the relative motion of the initial state and observer.

Lest one think that the QG propagator from~\eqref{eq:QGP} will be somewhat more complicated than~\eqref{eq:flat propagator}, note first of all that since we are dealing with a globally flat spacetime, the propagators for parallel transport are path independent and so one would not expect there to be a more general propagator built from it that would give a different result.
However, explicitly considering the integral
\begin{subequations}
\begin{equation}\label{eq:flat reproduce}
\int_{\sigma_t\times V_m^+}d\Sigma_m(x,p)
K\bigl(\tilde{\zeta}(x_2);
X_M(x)-i\tfrac{l}{m}\Lambda_O(x)p\bigr)
K\bigl(X_M(x)-i\tfrac{l}{m}\Lambda_O(x)p;
\tilde{\zeta}(x_1)\bigr),
\end{equation}
and transforming the momentum variables as $p\rightarrow \Lambda_O(x)^{-1}p$ we find
\begin{equation}\label{eq:flat reproduce 2}
\int_{\sigma_t} 2p\cdot\Lambda_O(x)d\sigma(x)\int_{V_m^+}d\Omega_m(p)
K\bigl(\tilde{\zeta}(x_2);
X_M(x)-i\tfrac{l}{m}p\bigr)
K\bigl(X_M(x)-i\tfrac{l}{m}p;
\tilde{\zeta}(x_1)\bigr).
\end{equation}
\end{subequations}
Now we see that the argument $X_M(x)-i\tfrac{l}{m}p$ is precisely the Minkowski coordinates evaluated in the parallelly propagated frame $e_A^O(x)$, and furthermore the boosted measure $\Lambda_O(x)d\sigma(x)$ is just the surface measure evaluated in these same hyper-plane frames (the surface measure transforms as a vector under changes of frame~\cite[Appendix B]{Wald:1984}).
By assumption the chosen coordinates $x$ constitutes a global chart (at least in some open region containing $\sigma$), and therefore $x\rightarrow X_M(x)$ is smooth and invertible.
The measure may therefore be written in terms of $X_M$, the result of which is an integral over $\sigma_t$ written in terms of standard Minkowski coordinates.
This is a special case of~\eqref{eq:repeater}, which may be used to complete the proof that the propagator~\eqref{eq:flat propagator} reproduces, and is therefore equivalent to the quantum--geometric propagator $\mathbf{K}^s$ defined by~\eqref{eq:QGP}.

\section{Propagation in a Milne Spacetime}
\label{sect:Milne}

We now turn to a specific illustration of the results of the last section, for simplicity choosing to work in a Milne universe which has constant time slices that are partial Cauchy surfaces (timelike geodesics are complete to the future).
This simple case gives further insight into the problems of the conventional local quantum field theory, since there exists in the literature inequivalent quantisations of a scalar field on hyperboloids of Minkowski space~\cite{Gromes+Rothe+Stech:1973,diSessa:1974,Sommerfield:1974}.
At present, there are authors who claim that the hyperboloids are inappropriate for imposing the canonical commutation relations and defining an inner product~\cite{Tanaka+Sasaki:1996}, whereas others~\cite{Arcuri+Svaiter+Svaiter:1994} claim that one must make an appropriate choice of orthonormal mode expansion.
In either case one must appeal to the behaviour of the propagator or orthonormal mode functions in the extension of the Milne universe to the Rindler wedge, even though physics in this region of spacetime should (in a truly local theory) have no effect on the causal evolution of a system in the Milne universe.
This is merely a reflection of the nonlocal nature of the vacuum of conventional local quantum field theory~\cite{Gibbons:1979}.

In the case of QG propagation similar issues arise.
Although we are dealing with a region of spacetime which is flat, since the hypersurfaces are asymptotically null the propagator does \textit{not} reproduce on them.
Indeed as we shall see, one cannot perform the sum in~\eqref{eq:QGP} in this case since the integrals are not convergent.
This presents no conceptual problem since this form of quantum propagation is fundamentally nonlocal, however it does emphasise the role of strictly spacelike hypersurfaces; indeed it is not clear that one could formulate QG propagation of initial data given on a null or partially null hypersurface.

\subsection{The Milne Universe}

The Milne universe that we will consider here is the region of Minkowski space (we will write the conventional Minkowski coordinates as $(t_M,x_M,\vec{y})$) that is in the causal future of the origin in the $t_M-x_M$ plane, while the remaining two spatial directions will be covered by the flat coordinate vector $\vec{y}$. 
Specifically, with the $(t,x,\vec{y})$ coordinate domain $t>0$ and $-\infty<x,\vec{y}<\infty$, the Milne (coordinate frame) metric will be given by
\begin{subequations}\label{eq:metric}
\begin{equation}
\mathrm{g}=\mathrm{diag}(1,-t^2,-1,-1),\quad
\mathrm{g}^{-1}=\mathrm{diag}(1,-t^{-2},-1,-1),
\end{equation}
so that $\sqrt{-\mathrm{g}}=t$ or $E:=\det({E^A}_\mu)=t$, depending on whether one works in the coordinate frame or the Lorentz frame $\{\partial_t,e_x,\partial_{\vec{y}}\}$ with dual coframe basis $(dt,\theta^x,d\vec{y})$ where
\begin{equation}
\theta^x:=t\,dx,\quad
e_x:=t^{-1}\partial_x.
\end{equation}
\end{subequations}
Since the metric has the form of a flat Euclidean metric in the $y-z$ plane, we will employ the vector symbol in that sector to simplify the notation.
Making use of the coordinate transformation between the adopted Milne coordinates and the standard Minkowski coordinates
\begin{equation}
t_M=t\cosh(x),\quad x_M=t\sinh(x),\quad
t=\sqrt{t_M^2-x_M^2},\quad x=\tanh^{-1}(x_M/t_M),
\end{equation}
shows that the metric~\eqref{eq:metric} covers the $t_M>0$, $-t_M<x_M<t_M$, and $-\infty<\vec{y}<\infty$ region of Minkowski space.
The chosen Lorentz frame may be related to the flat Minkowski coordinate frame by
\begin{equation}\label{eq:Mink boost}
\begin{bmatrix}\partial_t& e_x & \partial_{\vec{y}}\end{bmatrix}
=\begin{bmatrix}\partial_{t_M}&\partial_{x_M} & \partial_{\vec{y}}\end{bmatrix}
\begin{bmatrix}\cosh(x)&\sinh(x)&\vec{0}\\\sinh(x)&\cosh(x)&\vec{0}\\
\vec{0}&\vec{0}&\mathbf{1}\end{bmatrix}:=(\partial_{x_M^A})\Lambda(x)^{-1},
\end{equation}
which is the usual Lorentz frame transformation $e_A\rightarrow e_A^\prime=e_B{\Lambda^B}_A$, and implies that the Milne frames are related to the standard Minkowski frames via a boost along the $\hat{x}$ axis by $-x$, which has been defined as the Lorentz boost $\Lambda(x)^{-1}$.

\subsection{Geodesics}

It is a straightforward calculation to show that the only non--zero components of the Levi--Civita connection one--form~\eqref{eq:Gammas} are 
\begin{equation}\label{eq:Milne gammas}
{\omega^t}_x:=\Gamma^t_{xx}\theta^x
=dx=t^{-1}\theta^x
={\omega^x}_t
:=\Gamma^x_{xt}\theta^x,
\end{equation}
the geodesics of which are of course straight lines.
However, since they will not appear so in Milne coordinates, and to demonstrate how the calculation of the propagator proceeds, we will compute the geodesics explicitly.

The tangent to a geodesic will be written as $u^A=(u^t,u^x,\vec{u}):={E^A}_\mu\mathrm{d}_\tau[x^\mu]$ where $\mathrm{d}_\tau$ indicates the derivative with respect to the geodesic parameter $\tau$, so that $u^t=\mathrm{d}_\tau[t]$, $u^x=t\mathrm{d}_\tau[x]$, and $\vec{u}=\mathrm{d}_\tau[\vec{y}]$, and the geodesic equations are
\begin{equation}
\mathrm{d}_\tau[u^t]+(u^x)^2/t=0,\quad
\mathrm{d}_\tau[u^x]+u^tu^x/t=0,\quad
\mathrm{d}_\tau[\vec{u}]=0.
\end{equation}
This, combined with the normalisation $\mathrm{g}(u,u)=\delta_\epsilon$, results in
\begin{equation}\label{eq:vels}
u^t=\pm\sqrt{(u_0/t)^2+S},\quad
u^x=:u_0/t,\quad \vec{u}=:\vec{u}_0,
\end{equation}
where $S:=(\vec{u}_0)^2+\delta_\epsilon$, the $+$ and $-$ signs refer to forward and backward--pointing geodesics respectively, and 
\begin{equation}
\delta_\epsilon=\begin{cases}
+1&\text{for timelike geodesics}\\
0 &\text{for null geodesics}\\
-1&\text{for spacelike geodesics}\end{cases}.
\end{equation}
Using $u_t$, the geodesic equations may be reparameterised in terms of the time coordinate as
\begin{equation}\label{eq:geqt}
\mathrm{d}_t x=\pm \frac{u_0}{t^2\sqrt{(u_0/t)^2+S}},\quad 
\mathrm{d}_t \vec{y}=\pm\frac{\vec{u}_0}{\sqrt{(u_0/t)^2+S}}.
\end{equation}
Integrating these gives the geodesics
\begin{equation}\label{eq:geo int}
\begin{split}
\tau_1-\tau_0=&\pm\begin{cases}
\frac{t}{S}{\sqrt{(u_0/t)^2+S}}\Big\rvert^{t_1}_{t_0}
&\text{for $S\neq 0$}\\
\frac{t^2}{2\lvert u_0\rvert}\Big\rvert^{t_1}_{t_0}&\text{for $S=0$}
\end{cases},\\
x_1-x_0=&\mp\frac{u_0}{\lvert u_0\rvert}
\begin{cases}
\ln\Big(\lvert u_0\rvert/t+\sqrt{(u_0/t)^2+S}\Big)\Big\rvert^{t_1}_{t_0}
&\text{for $S\neq 0$}\\
\ln\big(1/t\big)\Big\rvert^{t_1}_{t_0}&\text{for $S=0$}
\end{cases},\\
\vec{y}_1-\vec{y}_0=&\pm\vec{u}_0(\tau_1-\tau_0).
\end{split}
\end{equation}

We have included the full set of solutions of the geodesic equations in order to discuss the following important issue.
If we consider an initially future--pointing spacelike geodesic emanating from $x_0$ (in the $\vec{y}=0$ plane) on the surface labelled by $t=t_0$ (which requires that $\lvert u_0\rvert>t_0$), then~\eqref{eq:vels} tells us that these geodesics continue to be forward--pointing until $t_{\mathrm{max}}:=\lvert u_0\rvert$.
At this point, $u^t=0$ and $\lvert u^x\rvert=1$, and we see that the geodesic is tangent to $\sigma_{t_{\mathrm{max}}}$, after which the geodesic then becomes past--pointing.
Using~\eqref{eq:geo int} we find the maximum spatial distance that the geodesic may `travel' before encountering some later surface $\sigma_{t_1}$ is found by choosing $\lvert u_0\rvert=t_1$, resulting in $\Delta x_{\mathrm{max}}=\ln\bigl(t_1/t+\sqrt{(t_1/t)^2-1}\bigr)$.
In order to connect points on $\sigma_{t_0}$ to points on $\sigma_{t_1}$ with $\Delta x>\Delta x_{\mathrm{max}}$, one has to consider geodesics that pass through $\sigma_{t_1}$ as a forward--pointing geodesic, continue on until it tangents at $\sigma_{\lvert u_0\rvert}$ for some $\lvert u_0\rvert>t_1$ at which point it becomes past--pointing, and then continues on to reach $\sigma_{t_1}$ for a second time.
(This is straightforward to see if one considers the surfaces as hyperboloids in Minkowski space, and draws straight lines between arbitrary points on nearby hypersurfaces.)

A similar scenario arose in the consideration of parallel propagation in a Robertson--Walker spacetime~\cite{Coleman:1996}, however in that case the geodesic could be smoothly matched (at $\Delta x_{\mathrm{max}}$) to a geodesic that remained within the spatial hypersurface, and could therefore be used to connect all points with greater separation.
Here we will assume that some choice of path has been made to join such points in the construction of the QG--propagator~\eqref{eq:QGP}; since the spacetime is flat, all choices yield the same transition element.
This is justified \textit{post facto} by the agreement of the resulting propagator with that derived using the standard Minkowski space hyper-planes as the foliation of spacetime. 
However in a curved spacetime where parallel transport is path--dependent, the QG--propagator may have to be constructed with some care as to how to make such a choice.
In principle, this should follow from taking the semi-classical limit of quantum--geometric gravitational propagation given in~\cite[Chapter 8]{Prugovecki:1995}, although it is possible that a study of diffeomorphism invariance of a semi-classical model may shed some light on this issue.

\subsection{Parallel Transport}

Since we are interested in hypersurfaces of constant $t$, the geodesics may be reparameterised by $t$ (as in~\eqref{eq:geqt}) and all integrals in $\tau$ may be converted to integrals in $t$.
Using~\eqref{eq:Milne gammas}, the matrices appearing in~\eqref{eq:Mobius} (and the remainder of that section) are given by
\begin{equation}
\begin{split}
\tfrac{1}{2}{\omega^A}_B(u){{ }\tilde{M}^B}_A+\tilde{\theta}^A(u)\tilde{P}_A
=&\tfrac{1}{2}{\omega^A}_B(u){{ }\tilde{M}^B}_A+(u^A+\nabla_u[a]^A)\tilde{P}_A\\
=&u^xt^{-1}{{ }\tilde{M}^1}_0+(u^A+\nabla_u[a]^A)\tilde{P}_A.
\end{split}
\end{equation}

Considering any geodesic with endpoints relating the value of the time coordinate to the value of the $x$ coordinate as for example $x_1:=x(t_1)$, from~\eqref{eq:geodesic exponential} one finds that the exponential determining the Lorentz transformation has as it's argument 
\begin{subequations}
\begin{equation}
-\int_{\tau_1}^{\tau_2}\frac{u^x}{t} d\tau
=-\int_{t_1}^{t_2}\frac{dx}{dt} dt=-(x_2-x_1),
\end{equation}
and so the Lorentz boost is given by
\begin{equation}\label{eq:boost}
\Lambda(t_2,t_1):=
\exp\Bigl(-\int_{\tau_1}^{\tau_2}\frac{u^x}{t}d\tau K_1\Bigr)
=\Lambda(x_2)\Lambda(x_1)^{-1},
\end{equation}
\end{subequations}
where $\Lambda(x)$ was defined in~\eqref{eq:Mink boost}, and the hyperbolic angle addition formulae: $\cosh(a+b)=\cosh(a) \cosh(b)+\sinh(a) \sinh(b)$ and $\sinh(a+b)=\cosh(a)\sinh(b)+\sinh(a)\cosh(b)$ have been used.

If one works in the Lorentz section $s_0=(e_A,0)$, the translational contribution is given by
\begin{equation}
\begin{split}
b(t_2,t_1)=&-\int^{\tau_2}_{\tau_1}d\tau\,\Lambda(\tau_2,\tau)u(\tau)
=-\Lambda(x_2)\int^{t_2}_{t_1}dt\,\Lambda\bigl(x(t)\bigr)u(t)/u^t(t)\\
=&-\Lambda(x_2)\int^{t_2}_{t_1}dt\,
\begin{bmatrix}\cosh\bigl(x(t)\bigr)&\sinh\bigl(x(t)\bigr)&\vec{0}\\
\sinh\bigl(x(t)\bigr)&\cosh\bigl(x(t)\bigr)&\vec{0}\\
\vec{0}&\vec{0}&\mathbf{1}\end{bmatrix}
\begin{bmatrix}1\\ t\partial_tx\\ \partial_t\vec{y} \end{bmatrix}.
\end{split}
\end{equation}
The contribution to the $y-z$ plane is straightforward to integrate, and the remainder is the two component vector in the $t-x$ plane
\begin{equation}
-\int_{t_1}^{t_2}dt\begin{bmatrix}
\cosh\bigl(x(t)\bigr)+t\partial_t[x]\sinh\bigl(x(t)\bigr)\\
\sinh\bigl(x(t)\bigr)+t\partial_t[x]\cosh\bigl(x(t)\bigr)
\end{bmatrix}.
\end{equation}
The second term in each of these is of the form $t\partial_t[\sinh\text{ or }\cosh]\bigl(x(t)\bigr)$ and may be integrated by parts, cancelling off the first term and leaving the surface contributions, giving the total contribution to the translation
\begin{subequations}\label{eq:translation}
\begin{equation}
b(t_2,t_1)=-\Lambda(x_2)
\begin{bmatrix}t_2\cosh(x_2)-t_1\cosh(x_1)\\t_2\sinh(x_2)-t_1\sinh(x_1)\\
(\vec{y}_2-\vec{y}_1)\end{bmatrix}
=\Lambda(x_2)\Lambda(x_1)^{-1}b(x_1)-b(x_2),
\end{equation}
where we have defined the vector
\begin{equation}
b(x):=(t,0,\vec{y}).
\end{equation}
\end{subequations}
Note that $\mathbf{b}(x):=b^Ae_A$ is the Minkowski coordinate vector $\mathbf{X}_M:=x_M^A\partial_{x_M^A}$ translated to this frame.
Using these results the Poincar\'{e} transformation~\eqref{eq:transition element} is given by
\begin{equation}
g_\gamma(t_2,t_1)=
\bigl(\Lambda(x_2,x_1),b(x_2,x_1)\bigr)=
\bigl(\Lambda(x_2),-b(x_2)\bigr)\star\bigl(\Lambda(x_1),-b(x_1)\bigr)^{-1}.
\end{equation}

Choosing instead the Poincar\'{e} frame $s_\parallel:=(e_A,\mathbf{a})$ with 
\begin{equation}\label{eq:a def}
a=-(t-t_0\cosh(x-x_0),t_0\sinh(x-x_0),\vec{y}-\vec{y}_0)
=\Lambda(x)\Lambda^{-1}_0 b_0-b(x),
\end{equation}
we find that in this case $b\equiv 0$.
Here the vector $-\mathbf{a}(x)$ may be seen to be the translation of the Minkowski coordinate vector $\mathbf{X}_M-(\mathbf{X}_M)_0:=(x_M^A-(x_M)^A_0)\partial_{x_M^A}$ to the Lorentz frame, where $(t_M)_0=t_0\cosh(x_0)$, $(x_M)_0=t_0\sinh(x_0)$, $\Lambda_0:=\Lambda(x_0)$ and $b_0:=b(x_0)$. 
Using~\eqref{eq:transition relation} these two transition elements may be related by
\begin{equation}
g_\gamma^{s_0}(x_2,x_1)=
\bigl(\mathbf{1},\mathbf{a}(x_2)\bigr)\star
g_\gamma^{s_\parallel}(x_2,x_1)\star
\bigl(\mathbf{1},\mathbf{a}(x_1)\bigr)^{-1}.
\end{equation}
In the case at hand, we have that $g_\gamma$ is a Lorentz boost in the Poincar\'{e} section $s_\parallel$ chosen with $a$ as in~\eqref{eq:a def}, and so transforming back to the Lorentz frame $s_0$ does not affect the Lorentz contribution~\eqref{eq:boost}, only introducing the translational factor
\begin{equation}
\begin{split}
b(t_2,t_1)=&a(t_2)-\Lambda(t_2,t_1)a(t_1),\\
=&\Lambda(x_2)\Lambda^{-1}_0 b_0-b(x_2)
-\Lambda(x_2)\Lambda(x_1)^{-1}\bigl(
\Lambda(x_1)\Lambda^{-1}_0 b_0-b(x_1)
\bigr)\\
=&\Lambda(x_2)\Lambda(x_1)^{-1} b(x_1)-b(x_2),
\end{split}
\end{equation}
which is the Lorentz frame result~\eqref{eq:translation}.

\subsection{The Quantum--Geometric Propagator}

The computation of the parallel transport elements allows one to write down the propagator for parallel transport corresponding to~\eqref{eq:gauge propagator} for each of the choices of section $s_0$ and $s_\parallel$, which turn out to give an identical (up to a constant factor that either cancels or is removed by choosing $(x_M)_0=0$) result
\begin{equation}\label{eq:Milne kernel}
\begin{split}
K^{s_0}_{\gamma(x_2,x_1)}\bigl(x_2,\hat{\zeta}(x_2);x_1,\hat{\zeta}(x_1)\bigr)
=&K\bigl(X_M(x_2)-i\tfrac{l}{m}\Lambda(x_2)^{-1}p_2;
X_M(x_1)-i\tfrac{l}{m}\Lambda(x_1)^{-1}p_1\bigr),
\end{split}
\end{equation}
where we have identified $X_M(x)=\Lambda(x)^{-1}b(x)$ as the components of the Minkowski coordinate vector (note that $\mathbf{b}(x):=\mathbf{x}_M$ but $X_M(x)=x_M$, that is, the first is the actual vector translated to the new frame whereas the second is the components of the vector transformed). 
Using the measure $d\Sigma_m(x,p)=2p^0t_0\,d^3x\,d\Omega_m(p)$, the integral analogous to~\eqref{eq:flat reproduce} is
\begin{equation}\label{eq:Milne reproduce}
\int_{\sigma_t}2p^0t_0\,dx\,d^2\vec{y}\int_{V_m^+}d\Omega_m(p)
K\bigl(\tilde{\zeta}(x_2);
X_M(x)-i\tfrac{l}{m}\Lambda(x)^{-1}p\bigr)
K\bigl(X_M(x)-i\tfrac{l}{m}\Lambda(x)^{-1}p;\tilde{\zeta}(x_1)\bigr)
\end{equation}
where we have identified $\tilde{\zeta}(x_n):=X_M(x_n)-i\tfrac{l}{m}\Lambda(x_n)^{-1}p_n$.
Following the argument at the end of Section~\ref{sect:QP}, we transform the momentum variables (this time $p\rightarrow\Lambda(x)p$) resulting in $p^0=p\cdot n(x)\rightarrow p\cdot\Lambda(x)^{-1}n(x)$ where $n(x):=(1,\vec{0})$ is the normal to the surface evaluated in the frame $e_A(x)$.
The vector $\Lambda(x)^{-1}n(x)$ may be written as $n_M(x_M)=(t_M/t,x_M/t,\vec{0})$, which are the vector components of the normal to $\sigma_t$ ($n_M^A:=\eta^{AB}\partial_{x_M^B}[t]$--which in this case is already normalised) evaluated in the Minkowski frame associated with the coordinates $t_M,x_M,\vec{y}$.
The surface integral may also be rewritten as $\int_{\sigma_{t_0}}t\,dx\,d^2\vec{y}\equiv \int_\mathbf{M}\delta(t-t_0)t\,dt\,dx\,d^2\vec{y}$, which is translated to Minkowski coordinates to give $\int_\mathbf{M}\delta\bigl(\sqrt{t_M^2-x_M^2}-t_0\bigr)dt_M\,dx_M\,d^2\vec{y}$.
Thus we have transformed the integral into one over a hyperboloid in Minkowski space, the surface contribution to which, as we noted following~\eqref{eq:repeater}, does \textit{not} vanish, and the kernel~\eqref{eq:Milne kernel} does not reproduce (in fact the integral~\eqref{eq:Milne reproduce} is divergent).

\section{Discussion}

We have shown how the construction of the free QG propagator on a classical background spacetime proceeds, and furthermore shown that it fails for the Milne universe that we considered.
The problem is that we have been using hypersurfaces that are only partial Cauchy surfaces of the spacetime, which are not sufficient due to the fundamentally non-local nature of QG propagation.
If one had a \textit{truly} local theory, then one would expect that the evolution of any system to the future of a Milne hypersurface would be completely determined by data on this surface.
Indeed, one may consider a system which has radiated away prior to $t_M=0$ (so that the Milne sector is flat).
In terms of \textit{any} foliation that consists of global Cauchy surfaces, the quantum--geometric propagator will not be exactly equivalent to that of a flat spacetime, even for points that are well within the flat region of spacetime.
It is clear that the construction in Section~\ref{sect:Milne} cannot see this, and the lack of complete information is reflected in the fact that the propagator sum~\eqref{eq:QGP} is ill--defined.
In general the foliation of spacetime \textit{must} consist of global Cauchy surfaces (and therefore spacetime must be globally hyperbolic) in order to derive a consistent result.

Although the actual construction of a quantum--geometric propagator in a curved spacetime has not been carried out as yet, the knowledge of its general form in a flat spacetime already has some remarkable features.
Not only is the flat spacetime propagator~\eqref{eq:flat propagator} well--defined for any choice of coordinates, Poincar\'{e} frame of reference, or foliation of spacetime into spacelike hypersurfaces, but the results of propagation viewed by different observers is reflected by physically reasonable boosts of the observed momentum spectrum of a state which takes the place of the Bogolubov transformations of conventional field theory.

Thus we are naturally lead to consider the observer dependence of conventional quantum field theory on curved spacetime.
In particular, the \textit{ex--nihilo} particle production in Rindler spacetimes and it's interpretation as the thermal radiation of a detector~\cite{Unruh+Wald:1984} may be assessed in the context of a theory that is consistently extendible in a straightforward manner to curved spacetimes.
Furthermore, the preliminary results~\cite{Prugovecki:1996,Coleman:1996} lead to the consideration of particle production in an expanding universe, as well as a study of the possibility of macroscopic acausal effects due to the presence of strong curvatures.

It is also interesting to look for nontrivial effects of a consistent quantum field theory on a curved background in cosmological scenarios by considering the back-reaction of the presence of the scalar field on the classical spacetime through the local average quantum stress--energy $\langle T_{AB}\rangle$.
Although the ultimate formulation of Quantum--Geometric Gravity~\cite{Prugovecki:1996} is expected to be diffeomorphism invariant, this type of (ultimately approximate) model may not be.
It would therefore be interesting to determine how this lack of diffeomorphism invariance manifests itself, and therefore determine the limitations of this type of semi--classical model.

\section*{Acknowledgements}

The author thanks the Natural Sciences and Engineering Research Council of Canada for a postdoctoral fellowship.

\bibliographystyle{amsplain}
%\bibliography{gengr,hamilton,fieldth,Prugovecki,me}

\begin{thebibliography}{10}

\bibitem{Ali:1985}
S.~Twareque Ali, \emph{Stochastic localization, quantum mechanics on phase
  space and quantum space--time}, Riv. Nuovo Cimento \textbf{8} (1985), no.~11,
  1--128.

\bibitem{Ali+Antoine+Gazeau:1993a}
S.~Twareque Ali, J.-P. Antoine, and J.-P. Gazeau, \emph{Continuous frames in
  {H}ilbert space}, Ann. Phys. (N.Y.) \textbf{222} (1993), 1--37.

\bibitem{Ali+Antoine+Gazeau:1993b}
\bysame, \emph{Relativistic quantum frames}, Ann. Phys. (N.Y.) \textbf{222}
  (1993), 38--88.

\bibitem{Arcuri+Svaiter+Svaiter:1994}
R.C. Arcuri, B.F. Svaiter, and N.F. Svaiter, \emph{Is the {M}ilne coordinate
  system a good one?}, Mod. Phys. Lett. A \textbf{9} (1994), 19--27.

\bibitem{Bogolubov+:1990}
N.N. Bogolubov, A.A. Logunov, A.I. Oksak, and I.T. Todorov, \emph{General
  principles of quantum field theory}, Kluwer Academic Publishers, Dordrecht,
  1990.

\bibitem{Bohr+Rosenfeld:1950}
N.~Bohr and L.~Rosenfeld, \emph{Field and charge measurements in quantum
  electrodynamics}, Phys. Rev. \textbf{78} (1950), 794--798.

\bibitem{Clayton:1997c}
M.A. Clayton, \emph{Canonical general relativity: Diffeomorphism constraints
  and spatial frame transformations}.

\bibitem{Clayton:1996b}
\bysame, \emph{The tetrad frame constraint algebra}, Class. Quantum Grav.
  \textbf{14} (1997), 1851--1864.

\bibitem{Coleman:1996}
James Coleman, \emph{On parallel transport in quantum bundles over
  {R}obertson--{W}alker spacetimes}, gr--qc/9605067, 1996.

\bibitem{DeWitt:1962}
Bryce~S. DeWitt, \emph{The quantization of geometry}, Gravitation: an
  introduction to current research (New York) (Louis Witten, ed.), John Wiley
  and Sons, 1962, pp.~266--381.

\bibitem{DeWitt:1975}
\bysame, \emph{Quantum field theory in curved spacetime}, Phys. Rep.
  \textbf{19} (1975), 295--357.

\bibitem{diSessa:1974}
A.~diSessa, \emph{Quantization on hyperboloids and full space--time field
  expansion}, J. Math. Phys. \textbf{15} (1974), 1892--1900.

\bibitem{Donoghue+Golowich+Holstein:1992}
John~F. Donoghue, Eugene Golowich, and Barry~R. Holstein, \emph{Dynamics of the
  standard model}, Cambridge University Press, Cambridge, 1992.

\bibitem{Drechsler:1982}
W.~Drechsler, \emph{{P}oincar\'{e} gauge field theory and gravitation}, Ann.
  Inst. Henri Poincar\'{e} \textbf{XXXVII} (1982), 155--184.

\bibitem{Drechsler+Tuckey:1996}
W.~Drechsler and Philip~A. Tuckey, \emph{On quantum and parallel transport in a
  {H}ilbert bundle over spacetime}, Class. Quantum Grav. \textbf{13} (1996),
  611--632.

\bibitem{Fernandez+Frolich+Sokal:1992}
Roberto Fern\'{a}ndez, J\"{u}rg Fr\"{o}lich, and Alan~D. Sokal, \emph{Random
  walks, critical phenomena, and triviality in quantum field theory},
  Springer--Verlag, New York, 1992.

\bibitem{Gibbons:1979}
G.W. Gibbons, \emph{Quantum field theory in curved spacetime}, General
  Relativity: An {E}instein centenary survey (Cambridge) (S.W. Hawking and
  W.~Israel, eds.), Cambridge University Press, 1979.

\bibitem{Glimm+Jaffe:1987}
James Glimm and Arthur Jaffe, \emph{Quantum physics: A functional integral
  point of view}, Springer--Verlag, New York, 1987.

\bibitem{Gradshteyn+Ryzhik:1980}
I.S. Gradshteyn and I.M. Ryzhik, \emph{Table of integrals, series and
  products}, corrected and enlarged ed., Academic Press, New York, 1980.

\bibitem{Gromes+Rothe+Stech:1973}
Dieter Gromes, Heinz~J. Rothe, and Berthold Stech, \emph{Field quantization on
  the surface ${X}^2=$constant}, Nucl. Phys. \textbf{B75} (1973), 313--332.

\bibitem{Hehl+:1995}
Freidrich~W. Hehl, J.~Dermott McCrea, Eckehard~W. Mielke, and Yuval Ne'eman,
  \emph{Metric affine gauge theory of gravity: Field equations, {N}oether
  identities, world spinors, and breaking of dilation invariance}, Phys. Rep.
  \textbf{258} (1995), 1--171.

\bibitem{Kinoshita:1990}
T.~Kinoshita, \emph{Quantum electrodynamics}, World Scientific, Singapore,
  1990.

\bibitem{Kobayashi+Nomizu:1963}
Shoshichi Kobayashi and Katsumi Nomizu, \emph{Foundations of differential
  geometry}, Interscience Tracts in Pure and Applied Mathematics, vol.~{I},
  John Wiley and Sons, New York, 1963.

\bibitem{Kuchar:1976a}
Karel Kucha\v{r}, \emph{Geometry of hyperspace. {I}}, J. Math. Phys.
  \textbf{17} (1976), 777--791.

\bibitem{Kuchar:1988}
Karel~V. Kucha\v{r}, \emph{the problem of time in canonical quantization of
  relativistic systems}, Conceptual Problems of Quantum Gravity (Boston) (Abhay
  Ashtekar and John Stachel, eds.), vol.~2, Birkh\"{a}user, 1988.

\bibitem{MTW:1973}
Charles~W. Misner, Kip~S. Thorne, and John~Archibald Wheeler,
  \emph{Gravitation}, W. H. Freeman and Company, New York, 1973.

\bibitem{Nakahara:1990}
M.~Nakahara, \emph{Geometry, topology and physics}, Adam Hilger, New York,
  1990.

\bibitem{Prugovecki:1984}
Eduard Prugove\v{c}ki, \emph{Stochastic quantum mechanics and quantum
  spacetime. {A} consistent unification of relativity and quantum theory based
  on stochastic spaces}, Fundamental Theories of Physics, D. Reidel Publishing
  Company, Holland, 1984.

\bibitem{Prugovecki:1992}
\bysame, \emph{Quantum geometry. {A} framework for quantum general relativity},
  Fundamental Theories of Physics, Kluwer Academic publishers, The Netherlands,
  1992.

\bibitem{Prugovecki:1994}
\bysame, \emph{Quantum--geometric field propagation in curved spacetime},
  Class. Quantum Grav. \textbf{11} (1994), 1981--1994.

\bibitem{Prugovecki:1995}
\bysame, \emph{Principles of quantum general relativity}, World Scientific,
  Singapore, 1995.

\bibitem{Prugovecki:1996}
\bysame, \emph{On quantum--geometric connections and propagators in curved
  spacetime}, Class. Quantum Grav. \textbf{13} (1996), 1007--1021.

\bibitem{Schweber:1961}
Silvan~S. Schweber, \emph{An introduction to relativistic quantum field
  theory}, Row, Peterson and Company, New York, 1961.

\bibitem{Sommerfield:1974}
Charles~M. Sommerfield, \emph{Quantization on spacetime hyperboloids}, Ann.
  Phys. (N.Y.) \textbf{84} (1974), 285--302.

\bibitem{Tanaka+Sasaki:1996}
Takahiro Tanaka and Misao Sasaki, \emph{Quantized gravitational waves in the
  {M}ilne universe}, gr-qc/9610060.

\bibitem{Teitelboim:1973}
Claudio Teitelboim, \emph{How commutators of constraints reflect the spacetime
  structure}, Ann. Phys. \textbf{79} (1973), 542--557.

\bibitem{Unruh+Wald:1984}
William~G. Unruh and Robert~M. Wald, \emph{What happens when an accelerating
  observer detects a {R}indler particle}, Phys. Rev. D \textbf{15} (1984),
  1047--1056.

\bibitem{Wald:1984}
Robert~M. Wald, \emph{General relativity}, The University of Chicago Press,
  Chicago, 1984.

\end{thebibliography}

\providecommand{\bysame}{\leavevmode\hbox to3em{\hrulefill}\thinspace}

\end{document}